\documentclass[a4paper,
               ]{jacow}
\usepackage{pdfpages,multirow,ragged2e} %
\usepackage{listings,xspace}
\makeatletter%
	\ifboolexpr{bool{xetex}}
	 {\renewcommand{\Gin@extensions}{.pdf,%
	                    .png,.jpg,.bmp,.pict,.tif,.psd,.mac,.sga,.tga,.gif,%
	                    .eps,.ps,%
	                    }}{}
\makeatother

\ifboolexpr{bool{xetex} or bool{luatex}} % test for XeTeX/LuaTeX
 {}                                      % input encoding is utf8 by default
 {\usepackage[utf8]{inputenc}}           % switch to utf8

\usepackage[USenglish]{babel}

\ifboolexpr{bool{jacowbiblatex}}%
 {%
  \addbibresource{jacow-test.bib}
  \addbibresource{biblatex-examples.bib}
 }{}
\lstdefinestyle{madngstyle}
{
  language       = {[5.2]Lua},
  deletekeywords = [2]{load},
  basicstyle     = \ttfamily\footnotesize,
  keywordstyle   = \color{magenta},
  commentstyle   = \color{orange},
  otherkeywords  = {..,:=,->,=>,\\,\#},
  emph           = {sequence,survey,track,cofind,twiss,taper,correct, errors,plot,match,normal,damap,
  jacobian,variables,equalities,command,beam},
  emphstyle      = \color{blue},
  emph           = {[2]nv,np,mo,po,vn,pn},
  emphstyle      = {[2]\color{red}},
  string         = [b]",
  stringstyle    = \color{brown},
}
\lstdefinelanguage{madx}
{
  basicstyle     = \ttfamily\footnotesize,
  keywords       = {},
  emph           = {normal,trackrdts},
  emph           = {[2]no,icase,exact,time},
  emphstyle      = {[2]\color{red}},
  string         = [b]",
  stringstyle    = \color{brown},
}

\newcommand{\MADX}{\lstinline{MADX}\xspace}

\begin{document}

\title{MAD-NG, a standalone multiplatform tool for \\ linear and non-linear optics design and optimisation}

\author{L.~Deniau\thanks{laurent.deniau@cern.ch}, CERN, Geneva, Switzerland \\[2ex]
{\normalsize Appears in the proceedings of the 14th International Computational Accelerator Physics Conference (ICAP’24)}\\
{\normalsize 2-5 October 2024, Germany}}

\maketitle

\begin{abstract}
The paper will provide an overview of the capabilities of the Methodical Accelerator Design Next Generation (MAD-NG) tool. MAD-NG is a standalone, all-in-one, multi-platform tool well-suited for linear and nonlinear optics design and optimization, and has already been used in large-scale studies such as HiLumi-LHC or FCC-ee. It embeds LuaJIT, an extremely fast tracing just-in-time compiler for the Lua programming language, delivering exceptional versatility and performance for the forefront of computational physics. The core of MAD-NG relies on the fast Generalized Truncated Power Series Algebra (GTPSA) library, which has been specially developed to handle many parameters and high-order differential algebra, including Lie map operators. This ecosystem offers powerful features for the analysis and optimization of linear and nonlinear optics, thanks to the fast parametric nonlinear normal forms and the versatile matching command. A few examples and results will complete this overview of the MAD-NG application.
\end{abstract}

\section{Overview}

MAD-NG (Methodical Accelerator Design Next Generation) \cite{MADNG-Home, MADNG-Git} is a modern stand-alone cross-platform download-and-run application developed for particle accelerator beam physics modeling, simulation, and optimization. As a successor to the MAD series codes, MAD-NG is built on decades of expertise and methodology in accelerator beam physics while offering advanced computational techniques, modularity, flexibility, scalability, and interoperability with external tools, tailored to meet the demands of contemporary accelerator complex designs.

MAD-NG natively supports lattice descriptions in MAD8\cite{MAD8}, MAD-X\cite{MADX, MADX-Home, MADX-Git}, and MAD-NG formats, ensuring seamless compatibility with legacy and modern models. It retains the full set of deferred expressions -- about a hundred thousand for the LHC (Large Hadron Collider, 27~Km) and HL-LHC (High Luminosity LHC upgrade) lattices -- to define functional knobs, circuits, and core lattice logic, preserving the design flexibility and functionality typical of the MAD series codes. Furthermore, MAD-NG allows users to work with multiple lattice descriptions simultaneously, making it ideal for analyzing entire accelerators complex or comparing different configurations of the same accelerator. For instance, it can efficiently handle and compare FCC-ee (Future Circular Collider, 91~Km) optics across various operational modes, such as the GHC (Global Hybrid Correction) and LCC (Local Chromatic Correction) designs at the Z and TTbar configurations.

MAD-NG features 5D and 6D PTC-like physics (Polymorphic Tracking Code)\cite{Forest02, Forest15, PTC-Git} considered as the gold standard for optics calculations. By leveraging high-order differential algebra and high-order symplectic integrators, it allows for seamless integration of various physics effects, including combined-function elements and user-defined physics. It also provides a comprehensive suite of high-level commands -- such as \lstinline{survey}, \lstinline{track}, \lstinline{twiss}, \lstinline{cofind}, \lstinline{match}, \lstinline{correct}, \lstinline{errors}, \lstinline{taper}, and \lstinline{plot} -- ensuring users familiar with MAD8, MAD9\cite{MAD9}, and MAD-X can transition to MAD-NG effortlessly. These commands maintain a consistent workflow in the same spirit as MAD codes, enabling users to conduct their studies with ease while taking advantage of MAD-NG’s enhanced capabilities.

MAD-NG supports forward, backward, and reverse tracking of charged particles and high-order differential maps, making it a great tool for working on the LHC dual beams and sequences, for example when squeezing beam optics for interaction point 1 (Atlas experiment at IP1) and 5 (CMS experiment at IP5) for both beams simultaneously. It also supports among others, thin and thick lens physics with various combination and integration models, sub-elements, elements fringe fields, exact misalignment and patches, radiation, weak-strong beam-beam, and flexible aperture shapes. 

MAD-NG offers high-order differential parametric map tracking alongside linear and non-linear parametric normal form analysis, enabling advanced studies and optimization of non-linear optics that would otherwise be unattainable. With performance speeds 50 to 80 times faster than MADX-PTC\cite{MADX-PTC}, it delivers the efficiency needed for demanding applications, such as training machine learning models or feeding online models with reasonable response time.

To avoid any confusion, the following acronyms will be used throughout this paper: MAD-X refers to the CERN code, MADX-PTC refers to E.~Forest's PTC/FPP library embedded into MAD-X, and the keyword \MADX refers to the special environment within MAD-NG that emulates the behavior of the global workspace of MAD-X. 

\section{MAD-NG Ecosystem}

\begin{figure}
\includegraphics[width=\linewidth]{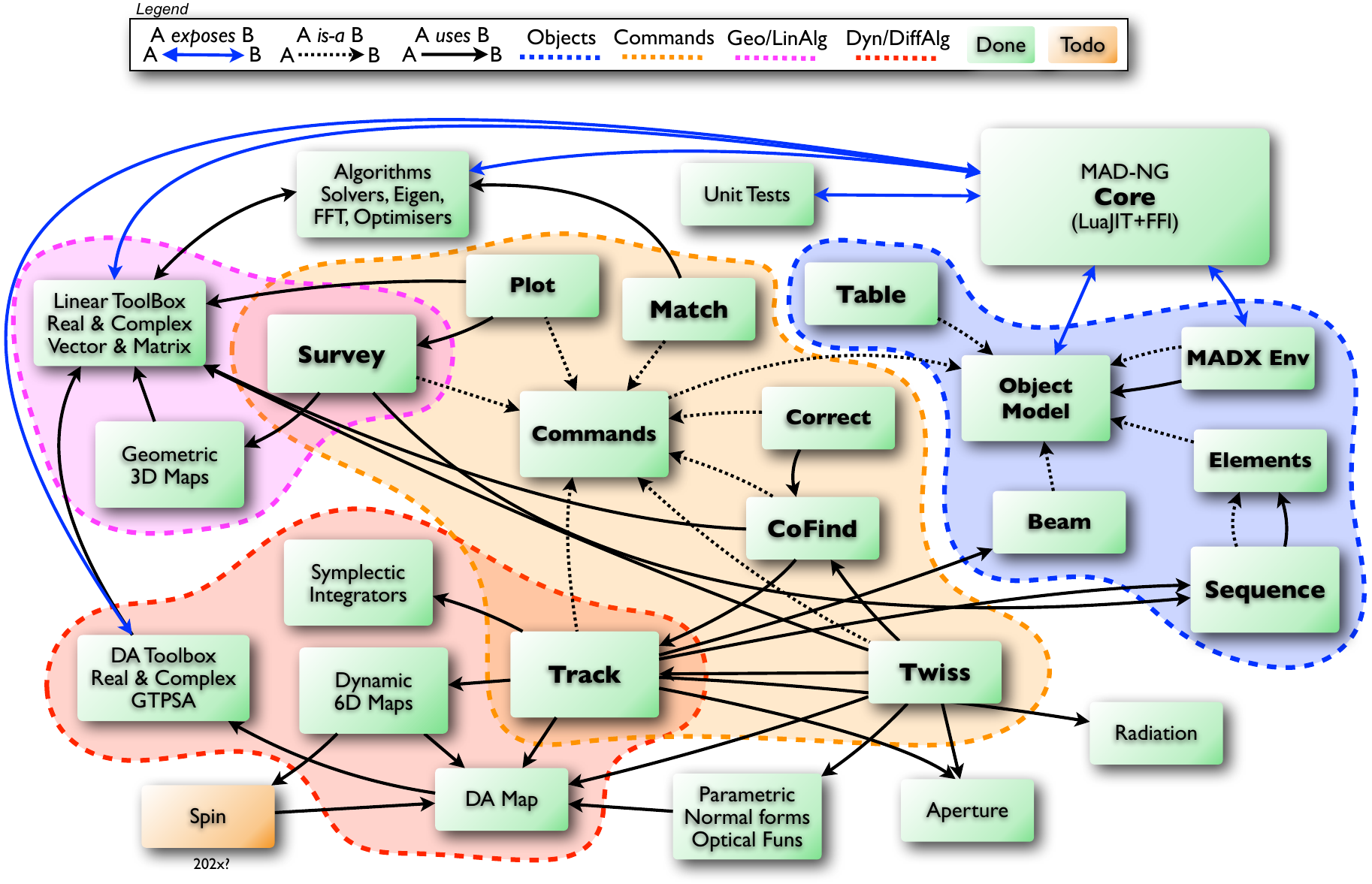}
\caption{MAD-NG ecosystem with all main components shown with relationship, and grouped by purpose: blue for objects, orange for commands, purple for 3D geometry and linear algebra, red for 6D dynamics and differential algebra.}
\label{ecosystem}
\end{figure}

\begin{figure}
\includegraphics[width=\linewidth]{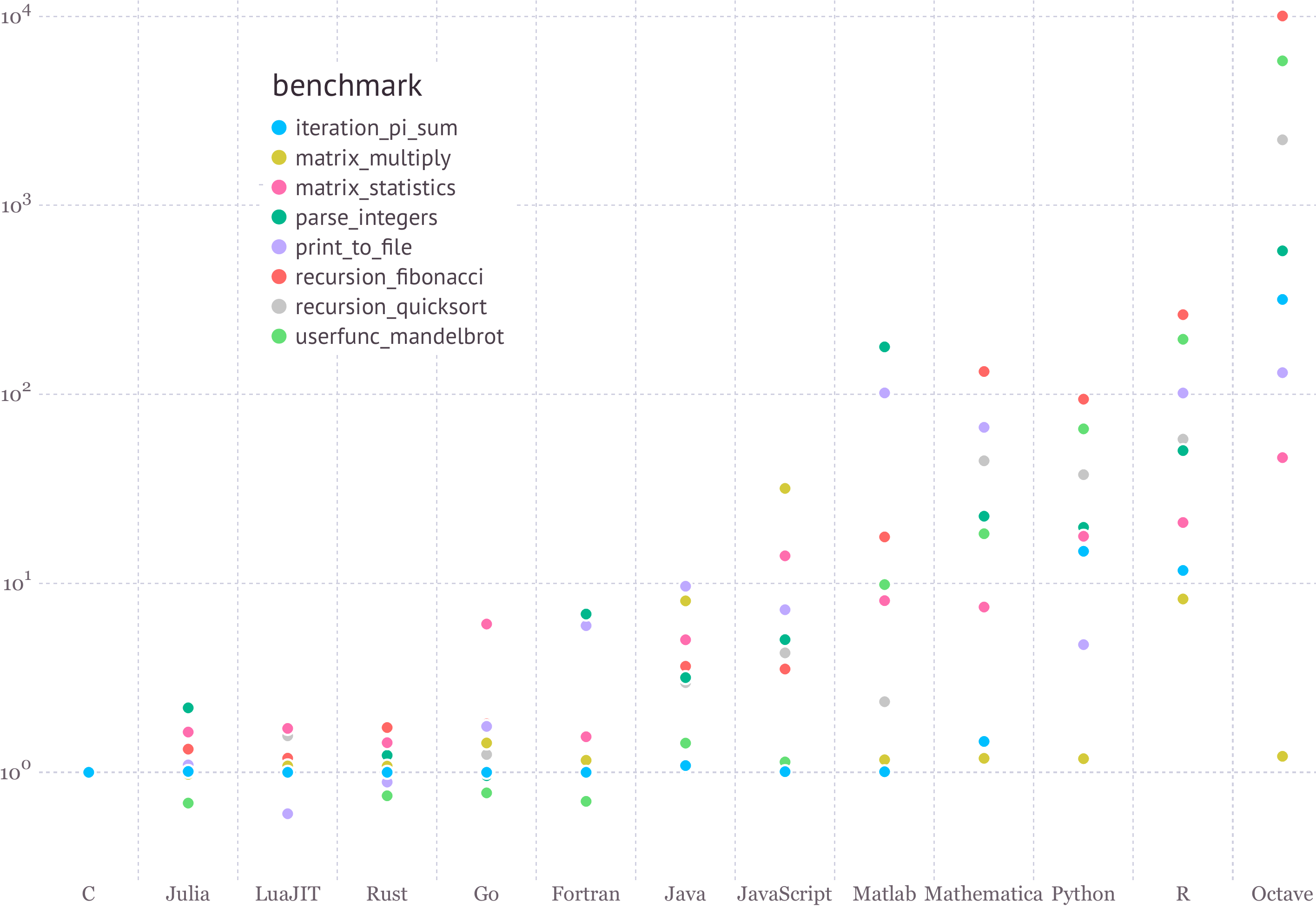}
\caption{The vertical axis shows benchmark times normalized to the C implementation, and LuaJIT in the 3rd column is one of the best performers along with Julia and Rust, while being the only one to be a dynamically typed programming language allowing rapid development.}
\label{luajit-perf}
\end{figure}

The MAD-NG scripting language is built on top of Lua\cite{Lua}, a widely adopted embedded programming language used across various industries to enhance application functionality. To maximize performance, MAD-NG incorporates the LuaJIT\cite{LuaJIT} tracing Just-In-Time compiler, one of the most advanced JIT compilers available. With approximately 70\% of MAD-NG’s code written in this scripting language, users gain full access to the core functionalities, modules, and libraries of the ecosystem schematically shown in Figure~\ref{ecosystem}, while achieving performance levels comparable to C/C++, as demonstrated by benchmarks in Figure~\ref{luajit-perf}.

\begin{figure}
\includegraphics[width=\linewidth]{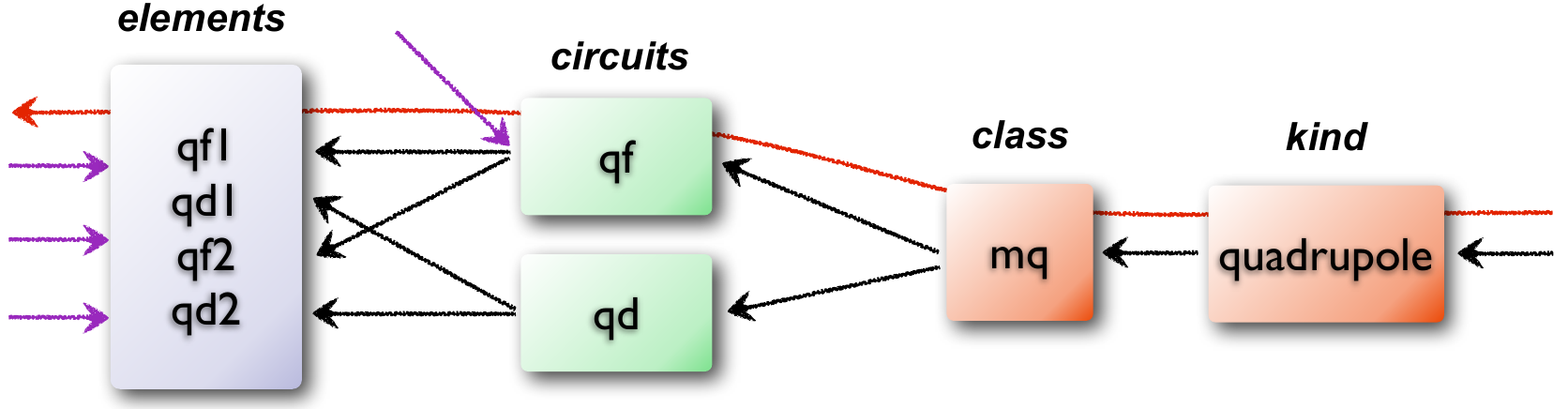}
\caption{The object model uses prototypal inheritance (black arrows) with dynamic lookup (red arrows) that stops at the first match, with intermediate attribute writes (purple arrows) shortening the search path.}
\label{object-model}
\end{figure}

The object model in MAD-NG (blue area in Figure~\ref{ecosystem}) introduces several essential features to maintain the flexibility that MAD users expect. One key aspect is the support for deferred expressions, familiar to MAD-X users as lazy expressions evaluated only when needed. MAD-NG extends this concept by generalizing deferred expressions into lambda functions, preserving their original semantics when used within the \MADX environment, which is itself treated as an object. Additionally, the object model includes dynamic inheritance, enabling efficient lookup within the object hierarchy to streamline functionality and enhance modularity, as shown in Figure~\ref{object-model}. Sequences, elements, beams, MAD tables (MTables) and commands are all objects that provide default attributes and behaviors to their children.

For example, reading, translating, and loading a complete and functional description of the LHC from a MAD-X file into MAD-NG's \MADX environment, retrieving the sequences \lstinline{lhcb1} and \lstinline{lhcb2} to assign them an injection proton beam takes a few hundred milliseconds on a standard laptop, and is as straightforward as:
\begin{lstlisting}
-- Load LHC sequences and optics setup
MADX:load "lhc_seq.madx"  
MADX:load "inj_optics.madx"
-- Create a proton beam at 450 GeV 
pb450 = beam { particle="proton", energy=450 }
-- Assign a beam to the two LHC sequences
local lhcb1, lhcb2 in MADX
lhcb1.beam, lhcb2.beam = pb450, pb450
 -- Tell that LHC B2 is a reversed sequence
lhcb2.dir = -1
\end{lstlisting}
The lattice descriptions loaded from the first file include thousands of deferred expression definitions, such as magnet and circuit strengths, which are configured by the second optics file loaded right after. Thanks to lazy evaluation and the \MADX environment, any variables that are referenced in expressions but not yet defined are automatically created on demand and initialized to zero, replicating the behavior of MAD-X itself. This eliminates the need for the user to manually manage variables and expressions bookkeeping or anticipate if an element's attribute will be a value or an expression, both being extremely error-prone at a such large scale. This flexibility lets users refine, modify, or switch optics setup simply by loading another file.

Finally, the following script snippet demonstrates the simplicity of working with different lattice configurations, as previously discussed, while avoiding name collisions between definitions:
\begin{lstlisting}
GHCZ = MADX {} :load "fcc_ghc_z.madx"  
GHCT = MADX {} :load "fcc_ghc_t.madx"  
LCCZ = MADX {} :load "fcc_lcc_z.madx"  
LCCT = MADX {} :load "fcc_lcc_t.madx"  
\end{lstlisting}
Each line creates a new distinct instance of the \MADX environment with an appropriate name, and fully equipped like the main instance through inheritance. It then loads the corresponding FCC-ee design, whether GHC or LCC in the Z or TTbar configuration, proceeding directly from the MAD-X files. These environments can be used conjointly or separately, as their memory footprint is minimal, typically much less then a hundred kilobytes. Accessing and modifying their content is just as straightforward as working with the main \MADX environment.

For Python users, the simple and intuitive interface enables transparent communication with MAD-NG subprocesses. Combined with PyMAD-NG\cite{PyMADNG}, it allows seamless script execution and efficient data retrieval in convenient formats, as illustrated in the Python script below:
\begin{lstlisting}
from pymadng import MAD
#-- Create an instance of MAD-NG
madng = MAD()
#-- Send script to MAD-NG
madng.send('''
  -- Copy here commands loading LHC as before
  local tw = twiss { sequence=lhcb1 }
  -- Send some table columns to Python
  pymad:send{tw.s, tw.beta11, tw.beta22}
  -- Send entire MTable to Python
  pymad:send(tw)
''')
#-- Receive MAD-NG vectors as Numpy arrays
s, beta11, beta22 = madng.recv()
#-- Receive and convert table to Pandas DF
tw = madng.recv().to_df()
\end{lstlisting}
The Python script begins by creating a MAD-NG instance as a sub-process with bidirectional communication via anonymous pipes. It then sends a complete MAD-NG script to the subprocess and retrieves the data returned as the result of script execution running a \lstinline{twiss} command. The data is sent in two ways for demonstration purposes: firstly, some data vectors of the Twiss table, then the entire Twiss MTable converted into a Pandas data frame.

\section{Sequences, Elements and Beams}

The \lstinline{sequence} command in MAD-NG defines the hierarchical layout of an accelerator as an ordered list of elements, specifying spatial arrangement, alignment, and connectivity of various components, such as magnets, drift spaces, RF cavities, monitors, etc. It supports nested subsequences for complex layouts, such as colliders or synchrotron rings, and chained sequences, such as racetracks or recirculators, enabling logical and modular modeling of entire accelerators complex.

When defining a sequence, users set key parameters like the starting point, element positions, and orientations. Its flexibility allows seamless adjustments to the lattice design, enabling direct exploration of element properties and the impact of different configurations, making it a powerful tool for design and optimization.

In MAD-NG, elements serve as the fundamental building blocks of an accelerator, representing specific physical or functional components like quadrupole magnets, dipole magnets, or drift spaces. Each element is defined with detailed attributes, including type, length, strength, alignment errors, and advanced features such as combined maps, custom apertures, combined fringe fields, and subelements for precise real-world modeling.

MAD-NG offers a diverse range of predefined element types, including sbends, rbends, true rbends, quadrupoles, sextupoles, octupoles, decapoles, dodecapoles, solenoids, kickers, monitors, drifts, markers, instruments, cavities, patches, electrostatic separators, compensating wires, beam-beam interactions, and more. It also supports generic linear and non-linear maps to model entire accelerator sections or specific localized physics. Furthermore, users can create custom elements or adapt existing ones to meet unique design requirements as the entire framework is written in scripting language and therefore accessible to users. 

The following example shows a concise description of the CERN SPS ring with a circumference of 7~Km from its early design phases with only 1128 elements, where operator overloading is used to handle repetitive elements and groups:
\begin{lstlisting}
pf     = bline {qf,2*b1,2*b2,ds}
pd     = bline {qd,2*b2,2*b1,ds}
p24    = bline {qf,dm,2*b2,ds,pd}
p42    = bline {pf,qd,2*b2,dm,ds}
p00    = bline {qf,dl,qd,dl}
p44    = bline {pf,pd}
insert = bline {p24,2*p00,p42}
super  = bline {7*p44,insert,7*p44}
SPS    = sequence "SPS" {6*super}
\end{lstlisting}

Once a sequence is defined with elements, users can employ commands like \lstinline{survey}, \lstinline{twiss}, \lstinline{track}, or \lstinline{match} to validate the lattice geometry, compute optical functions, simulate particle trajectories, or optimize lattice parameters. The modularity of sequences and elements ensures that modifications to one part of the accelerator design can be implemented safely using ranges or selectors without disrupting the entire model.

\section{Survey, Track, and Twiss}

The \lstinline{survey} and \lstinline{track} commands are critical tools for analyzing and optimizing particle accelerator designs. These commands facilitate the visualization and simulation of beam trajectories and the geometric layout of accelerators, providing insights into beam dynamics and the physical configuration of the system.

The \lstinline{survey} command is used to calculate and display the spatial geometry of an accelerator. It determines the global positions and orientations of accelerator elements along the beamline, accounting for offsets, tilts, rotations, and user-defined patches. This command is particularly useful for visualizing how individual components are arranged in three-dimensional space or geographic coordinate system, ensuring proper alignment and installation. The output typically includes Cartesian coordinates (X, Y, Z) and angular information (pitch, yaw, roll) for each element. Often used during the design phase, the \lstinline{survey} command facilitates communication between engineering and design teams and serves as a tool for validating layout accuracy after alignment updates.

\begin{figure}
\includegraphics[width=\linewidth]{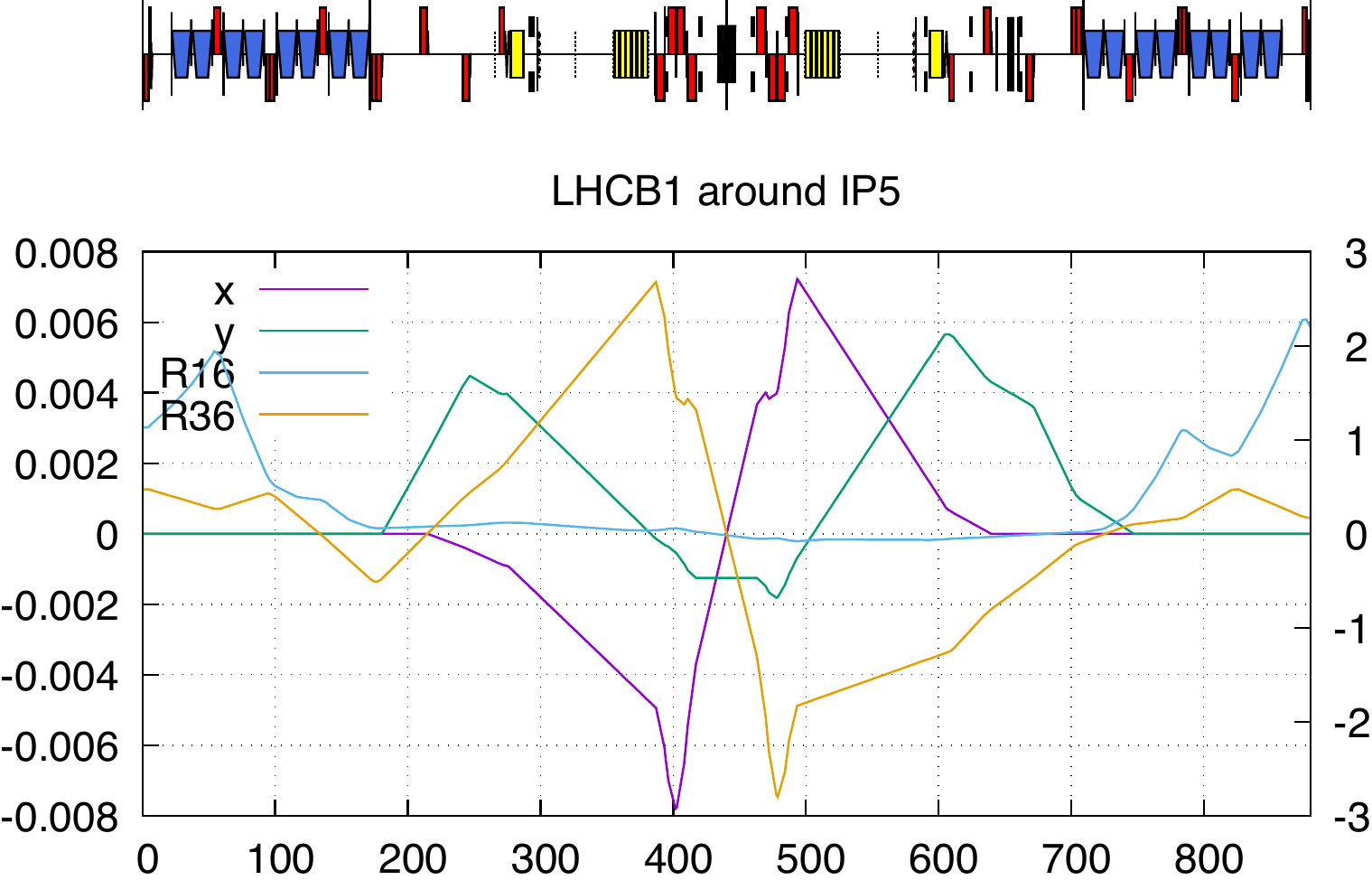}
\caption{Example of \lstinline{plot} using \lstinline{survey} to build the lattice layout (top), and \lstinline{track} to calculate the particle coordinates $x, y$ (left axis) and first derivatives $\frac{\partial x}{\partial p_t}=R_{16}, \frac{\partial y}{\partial p_t}=R_{36}$ (right axis) along $s$ around LHC IP5 (bottom axis). All axes are in meters.}
\label{track-ip5}
\end{figure}

The \lstinline{track} command is a simulation tool used to model the trajectories of particles through the accelerator lattice. By solving the equations of motion for charged particles under the influence of electromagnetic fields, this command predicts how beams propagate through the system. Users can specify beam characteristics like particle type and energy, as well as particles' initial position, momentum, and energy deviation, to evaluate how the accelerator handles different configurations. The output typically includes phase-space coordinates (e.g., position and momentum) at various points along the lattice as shown in Figure~\ref{track-ip5}, allowing for in-depth analysis of beam dynamics.

\begin{figure}
\includegraphics[width=\linewidth]{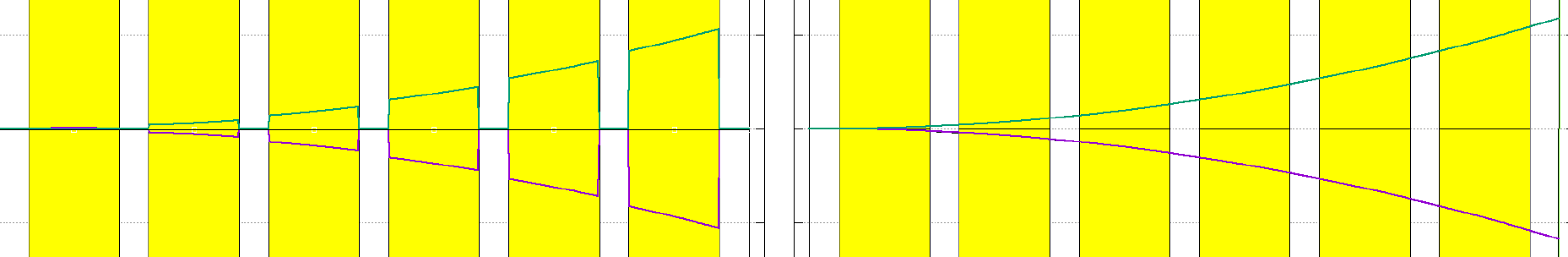}
\caption{Example using \lstinline{survey} and \lstinline{track} results to reconstruct the horizontal motion of the reference particle for the upstream (purple) and downstream (green) beams in the local frame $(x, s)$ on the left, and in the global frame $(X,Z)$ on the right.}
\label{track-d1}
\end{figure}

Figure~\ref{track-d1} illustrates an example that combines the \lstinline{survey} and \lstinline{track} command results to analyze the continuous motion of the reference particle in the global frame (right). In contrast, the motion appears discontinuous in the local frame (left) due to the automatic patches applied by the six true rbend magnets that constitute the LHC D1 responsible for bending the upstream and downstream beams around the interaction points.

\begin{figure}
\includegraphics[width=0.524\linewidth]{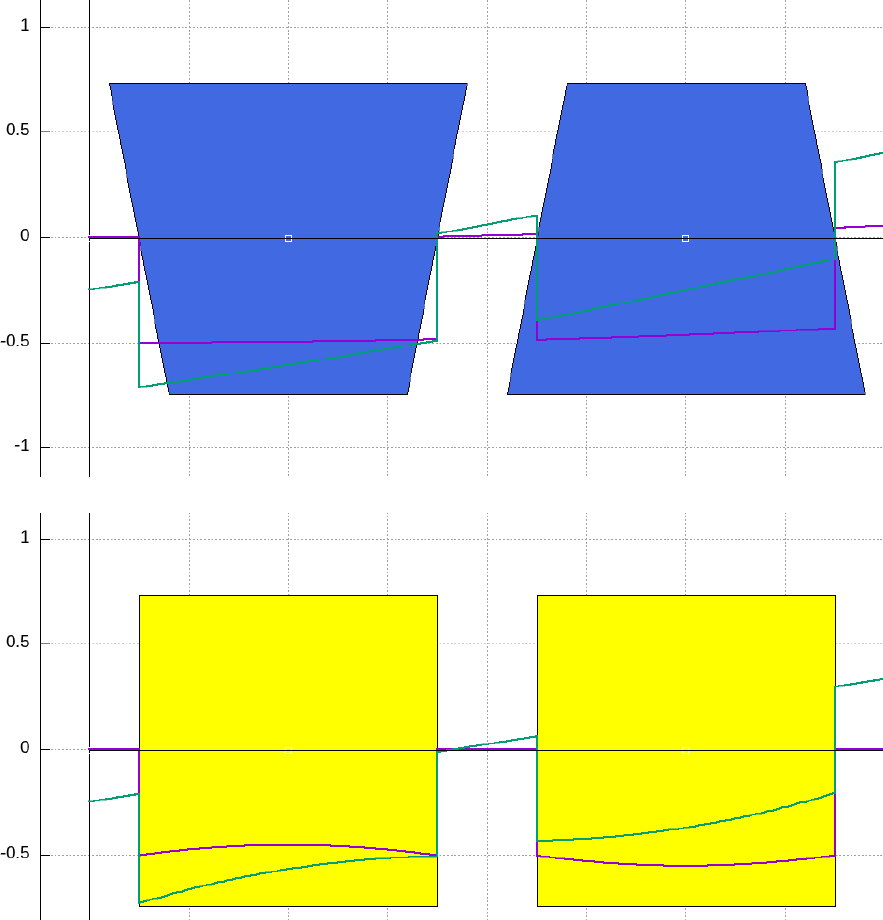}
\includegraphics[width=0.467\linewidth]{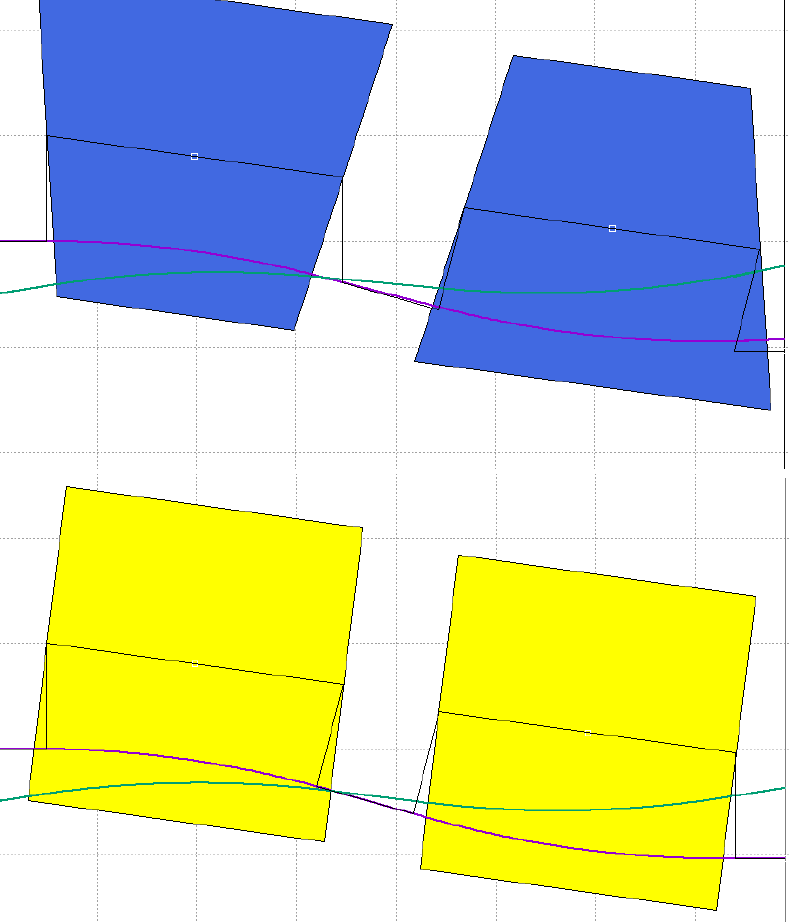}
\caption{Example using \lstinline{survey} and \lstinline{track} results to plot the horizontal motion of two particles (green, purple) with different initial coordinates vs the reference trajectory (black) in the local frame $(x, s)$ on the left, and in the global frame $(X,Z)$ on the right, while moving through a couple of misaligned sbends (top, blue) and true rbends (bottom, yellow).}
\label{track-misalgn}
\end{figure}

Figure~\ref{track-misalgn} shows a similar analysis performed to examine the effect of misalignment on particle motion, which is expected to be continuous in the global frame (right). However, in the local frame (left), the motion appears discontinuous due to the automatic patches that manage the misalignment at the entry and exit of the elements to restore the design orbit.

\begin{figure}
\begin{center}
\includegraphics[width=0.8\linewidth]{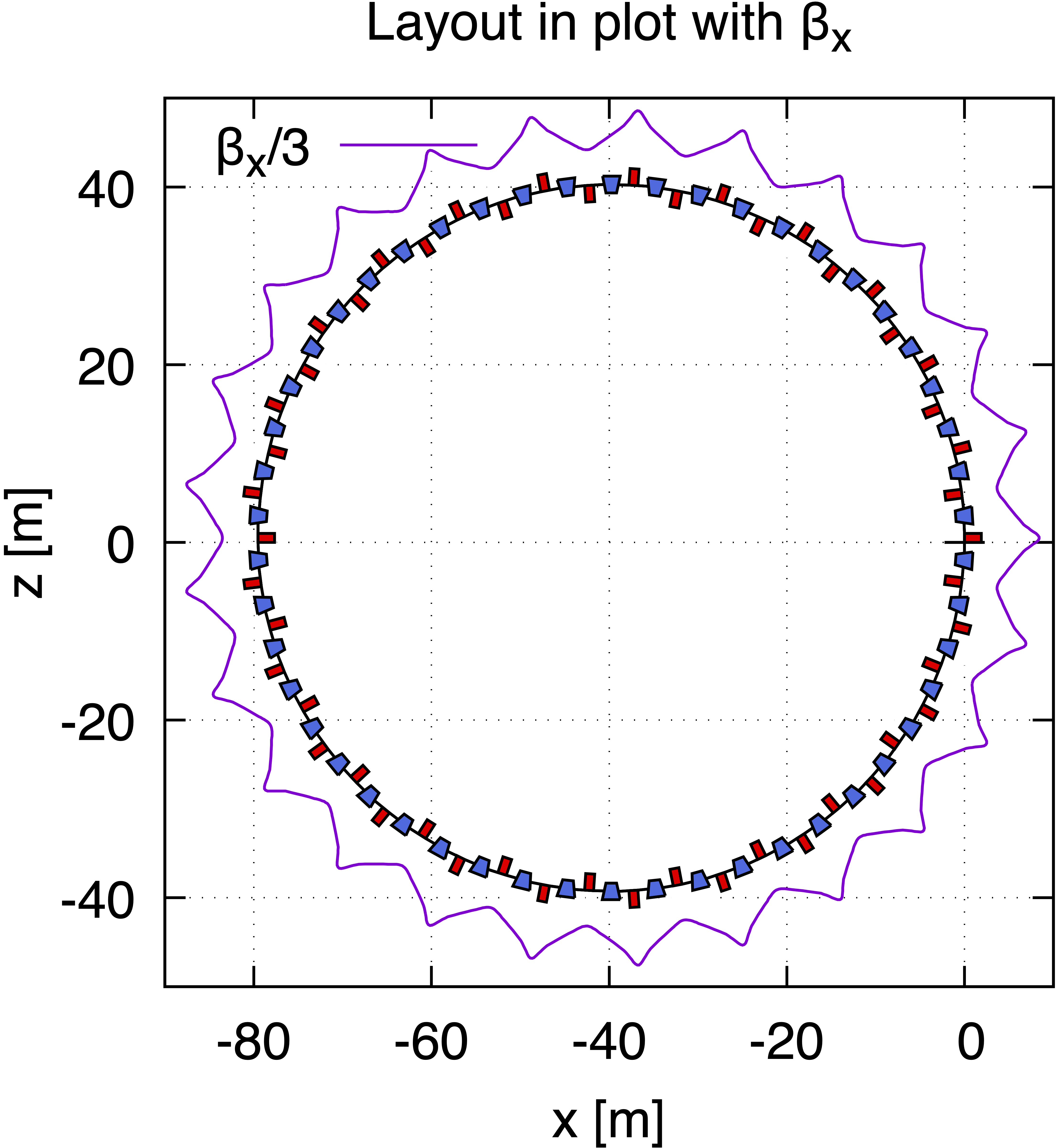}
\caption{Example using \lstinline{survey} and \lstinline{twiss} results to plot the scaled horizontal beta function $\beta_x$ in the global frame $(X,Z)$ on top of the layout made of 25 FODO cells.}
\label{layout-betx}
\end{center}
\end{figure}

Finally, the following concise MAD-NG script showcases the language’s expressiveness. In just a few lines, it constructs a toy lattice consisting of 25 FODO cells, calculates its geometrical and optical properties using the \lstinline{survey} and \lstinline{twiss} commands, and combines the results to extend the survey table, as illustrated in Figure~\ref{layout-betx}.
\begin{lstlisting}
-- Create the sequence from 25 fodo cells
local nc = 25
local mb = sbend { l=2, angle=pi/nc }
local mq = quadrupole { l=1 }
local cell = bline {
    mq "qf"  { at=0, k1= 0.29601 },
    mb "mb1" { at=2 },
    mq "qd"  { at=5, k1=-0.30242 },
    mb "mb2" { at=7 },
  }
local seq = sequence "ring" {
            nc*cell, refer="entry", beam=beam}
-- Run survey and save orientation matrix W 
local sv = survey {sequence=seq, mapsave=true}
-- Run twiss to get beta11
local tw = twiss { sequence=seq }
-- Compute beta x in global frame 
local X, Y, W = sv.x, sv.y, sv.W
local V, B = {}, tw.beta11/3+3
-- Build list of 3D oriented vectors
for i=1,#W do 
  V[i] = W[i] * vector{B[i],0,0}
end
-- Add column generators to survey table
sv:addcol('betx_X', \i -> V[i][1]+X[i])
sv:addcol('betx_Z', \i -> V[i][3]+Z[i])
-- Run plot with data picked from sv table 
\end{lstlisting}
Here, most calculations involve vector-matrix operations, leveraging operator overloading and features from the linear algebra module. This is because table columns storing only scalar values -- whether real or complex numbers -- are treated as vectors like in tables \lstinline{sv} and \lstinline{tw} returned by \lstinline{survey} and \lstinline{twiss} commands.

\begin{figure}
\includegraphics[width=\linewidth]{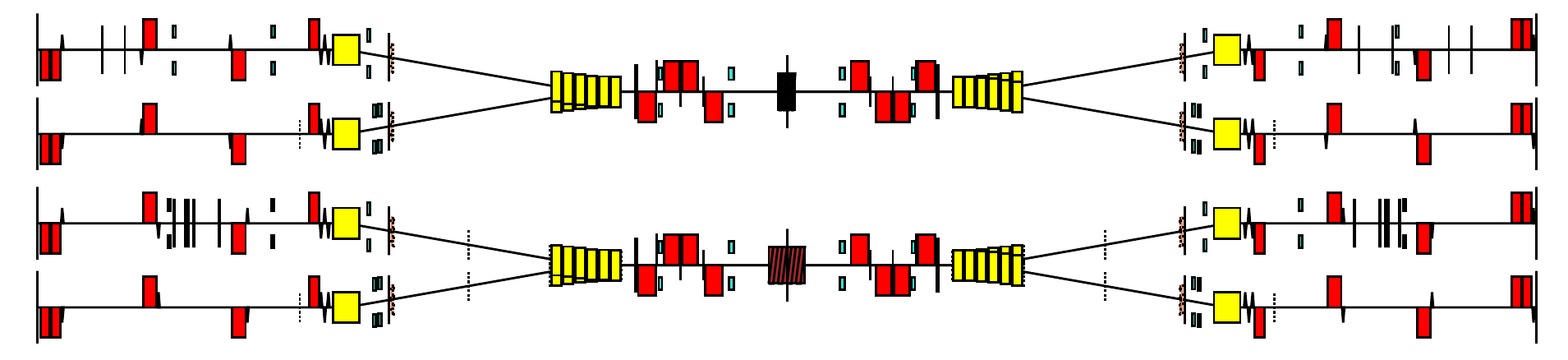}
\caption{Example drawing simultaneously four layouts, two beams around two IPs, and displayed by the \lstinline{plot} command above the data frame, leveraging \lstinline{survey} to position each element accurately.}
\label{layout-4seq}
\end{figure}

The \lstinline{plot} command itself relies on the \lstinline{survey} command to draw complex layouts like in Figure~\ref{layout-4seq} depicting LHC layout as installed for beams 1 \& 2 around IP 1 \& 5 with minimal setup:
\begin{lstlisting}
plot {
  sequence = { lhcb1, lhcb2, lhcb1, lhcb2 },
  range = { -- Ranges for the 4 sequences
    {"E.DS.L1.B1","S.DS.R1.B1"},-- B1 IP1 mark
    {"E.DS.L1.B2","S.DS.R1.B2"},-- B2 IP1  "
    {"E.DS.L5.B1","S.DS.R5.B1"},-- B1 IP5  "
    {"E.DS.L5.B2","S.DS.R5.B2"},-- B2 IP5  "
  },
  laydisty = {
    -- Second bline y-shift [m]
    lhcb2["E.DS.L1.B2"].mech_sep,       
    -- Third  bline y-shift [m]
    -0.4,                               
    -- Fourth bline y-shift [m]
    -0.4 + lhcb2["E.DS.L5.B2"].mech_sep
  },
}
\end{lstlisting}

In complement to the previous examples, the tracking results also provide insight into issues such as beam losses, instabilities, or deviations from the intended trajectory. This information is essential for optimizing element parameters, such as magnet strengths or RF field phases, to achieve the desired performance objectives.

It is worth noting that the \lstinline{survey} and \lstinline{track} commands serve as interfaces to the only highly configurable polymorphic tracking engines available in MAD-NG. The \lstinline{survey} command is dedicated to geometric tracking within the global frame, working with vectors and matrices, while the \lstinline{track} command performs dynamic tracking in the local mobile frame, operating polymorphically on either particle coordinates or high-order differential algebraic maps (DA maps). Other commands, such as \lstinline{taper}, \lstinline{cofind}, \lstinline{twiss}, or \lstinline{plot}, act as meta-commands that configure and run the \lstinline{survey} or \lstinline{track} commands for their specific purposes, often invoking them several times with different setups.

\section{Frames and Patches}

\begin{figure}
\includegraphics[width=\linewidth]{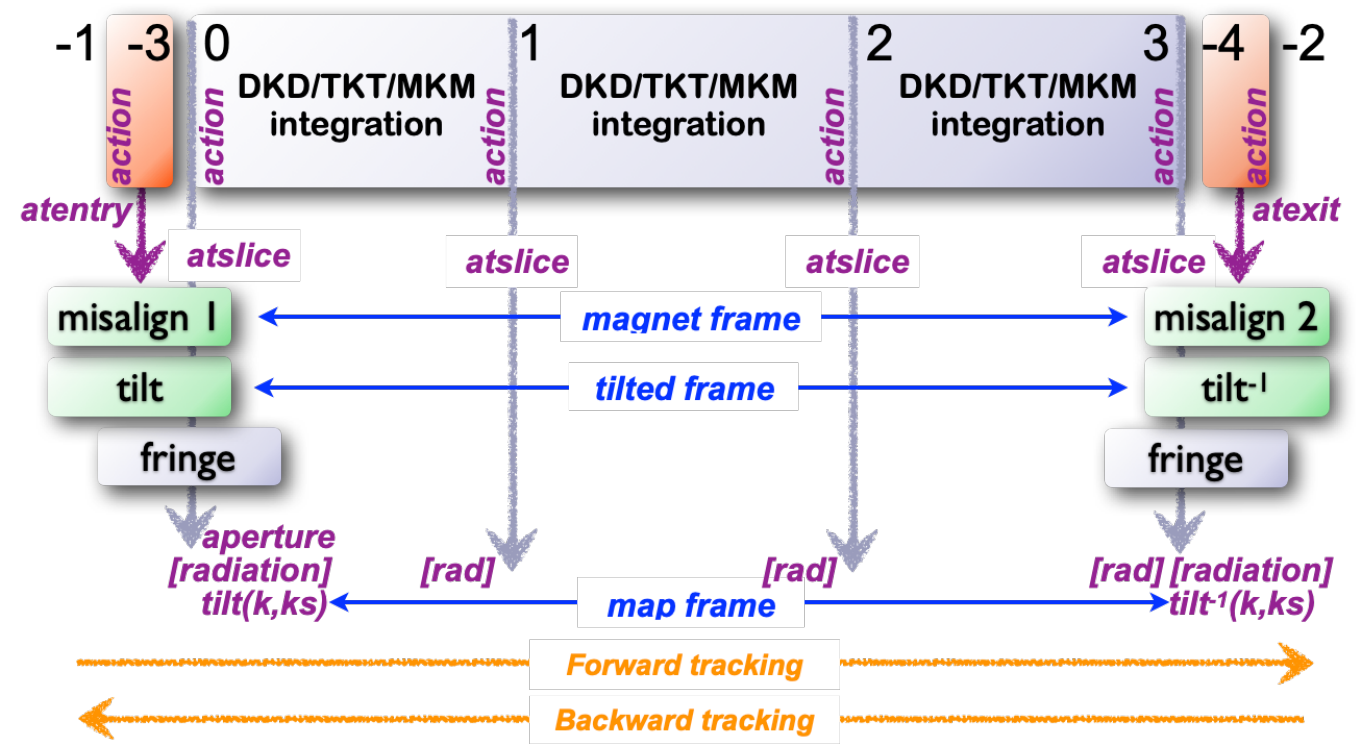}
\caption{Schematic behavior of the generic element tracker with stepwise frame transformations and slices integration.}
\label{elem-frames}
\end{figure}

The survey and the track engines run through sequence of elements. Upon entry, each element configures the generic element tracker with the appropriate functional maps to execute the most accurate physics based on its attributes' values. The element tracker looks like the following simplified sequence of function invocations: 
\begin{lstlisting}
  atentry(elm, mflw,  sdir)
  misalgn(elm, mflw,  sdir)
  tilt   (ang, mflw,  sdir)
  fringe (elm, mflw,  sdir) -- DKD/TKT/MKM
  integr (elm, mflw,    1, thick, kick)
  fringe (elm, mflw, -sdir) -- atslice
  tilt   (ang, mflw, -sdir)
  misalgn(elm, mflw, -sdir)
  atexit (elm, mflw, -sdir)
\end{lstlisting}
where the different steps can be visualized in Figure~\ref{elem-frames}. The \lstinline{thick, kick} are the functional maps selected by the element and invoked during each slice integration by the symplectic integrator \lstinline{integr}. The functional map \lstinline{fringe} is invoked on the boundaries, between the misalignment and tilt of the reference frame. The \lstinline{sdir} argument represents the forward or backward tracking direction, minuses on exit steps. An alternate more complex generic element tracker is used when sub-elements are present.

The special functions \lstinline{atentry}, \lstinline{atexit}, and \lstinline{atslice} execute all the actions -- typically lambda functions composed with slice selectors -- stored in their respective lists, whether they were set by users or other commands. This approach is the primary strategy employed by meta-commands like \lstinline{twiss} to configure the \lstinline{cofind} and \lstinline{track} commands during its multiple phases.

\begin{figure}
\begin{center}
\includegraphics[width=0.95\linewidth]{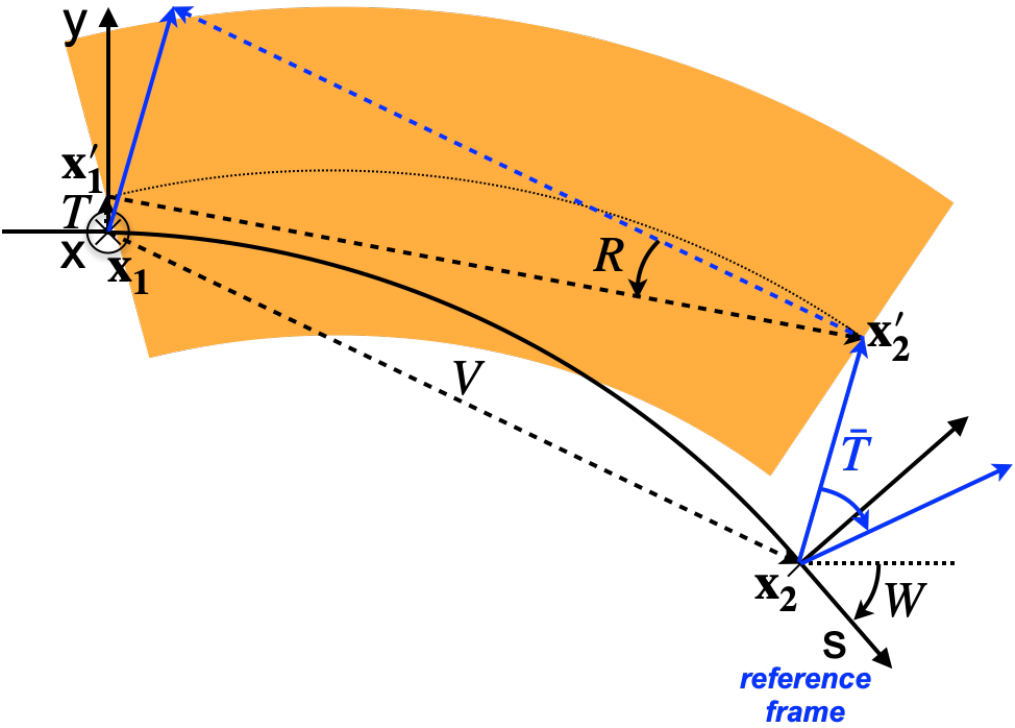}
\caption{Representation of misalignment of a sbend using computed patches to restore the curved reference frame on exit.}
\label{sbend-misalgn}
\end{center}
\end{figure}

Patches serve as fundamental components to handle misalignment and tilt of elements but with the added constraint of restoring the reference frame upon exit. This restoration is achieved by $\bar{T} = W^t(RV+T-V)$ and $\bar{R} = W^t R W$ given the vectors $T, V$ and the matrix $R$ as illustrated in Figure~\ref{sbend-misalgn}. The six true rbends modeling the LHC D1 magnet from Figure~\ref{track-d1} is another example of computed (implicit) patches used to restore the reference frame upon exit. In contrast, user-defined explicit patches introduce discontinuous frame transformations required, for example, to adjust the reference frame of a beam transfer line with that of an injection point in a circular accelerator.

\section{GTPSA and DA maps}

Truncated Power Series Algebra (TPSA), so named by M. Berz in the 80s, are truncated Taylor series based on automatic differentiation for calculating exact derivatives. They have proved to be extremely useful for solving differential equations when the problems and solutions can be modeled by real or complex analytic functions, as it is the case for many applications in physics. In 2014, I generalized the TPSA to GTPSA\cite{Deniau15} to support inhomogeneous orders and many parameters. As MAD-NG evolved, the library was improved and extended to support more operations like for example Lie operators and complex-valued functions. Nowadays, this library is regarded as the fastest, most versatile, and most advanced tool for handling high-order multivariate differential functions.

\begin{figure}
\includegraphics[width=\linewidth]{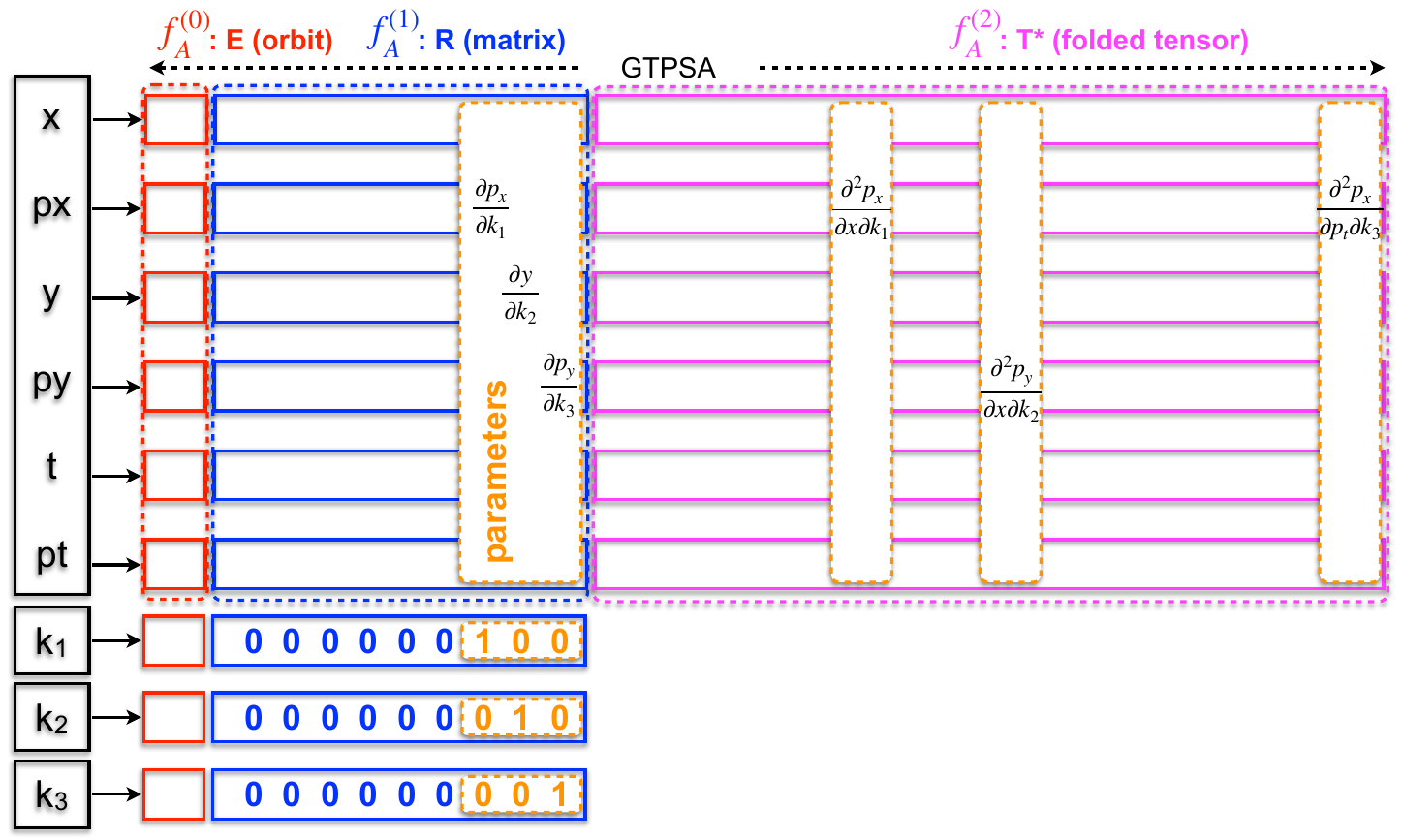}
\caption{Schematic representation of parametric DA maps with 6 variables ($x, p_x, y, p_y, t, p_t$) of order 2 and 3 parameters ($k_1, k_2, k_3$) of order 1, all made from GTPSA (row-wise).}
\label{gtpsa_layout}
\end{figure}

There are many methods to create differential algebra maps (DA maps) in MAD-NG. For instance the \lstinline{cofind} and the \lstinline{twiss} commands use DA maps with 6 variables of order 1 by default to compute the Jacobian matrix required by the closed orbit finder and the linear optics functions respectively. By specifying \lstinline{mapdef=2} to the \lstinline{twiss} command, the order is increased to 2, enabling the calculation of chromaticity. Higher orders and more customized DA maps definitions can be provided to the \lstinline{twiss} command to perform advanced non-linear analyses, as will be presented in the following.

The most common way to create standalone DA maps objects with 6 variables $x, p_x, y, p_y, t, p_t$ of order 4 and 3 parameters $k_1, k_2, k_3$ of order 1 is:
\begin{lstlisting}
local prms = {"k1", "k2", "k3"}
local X0 = damap {nv=6, mo=2,
                  np=#prms, po=1, pn=prms}
\end{lstlisting}
where \lstinline{nv} and \lstinline{np} specify the number of variables (default 6) and parameters (default 0) respectively, and \lstinline{mo} the maximum overall order of the GTPSA (default 1). The optional \lstinline{po} and \lstinline{pn} specify the parameters' maximum orders (default 1) and list of names.

\begin{figure}
\includegraphics[width=\linewidth]{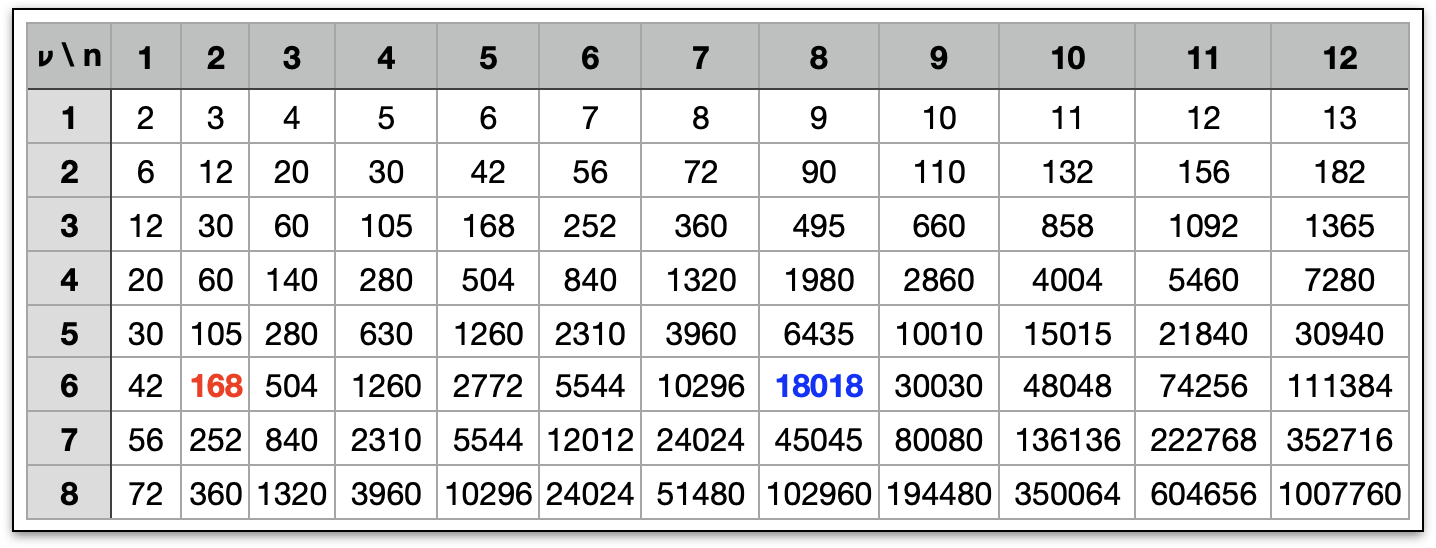} \\[1ex]
\includegraphics[width=\linewidth]{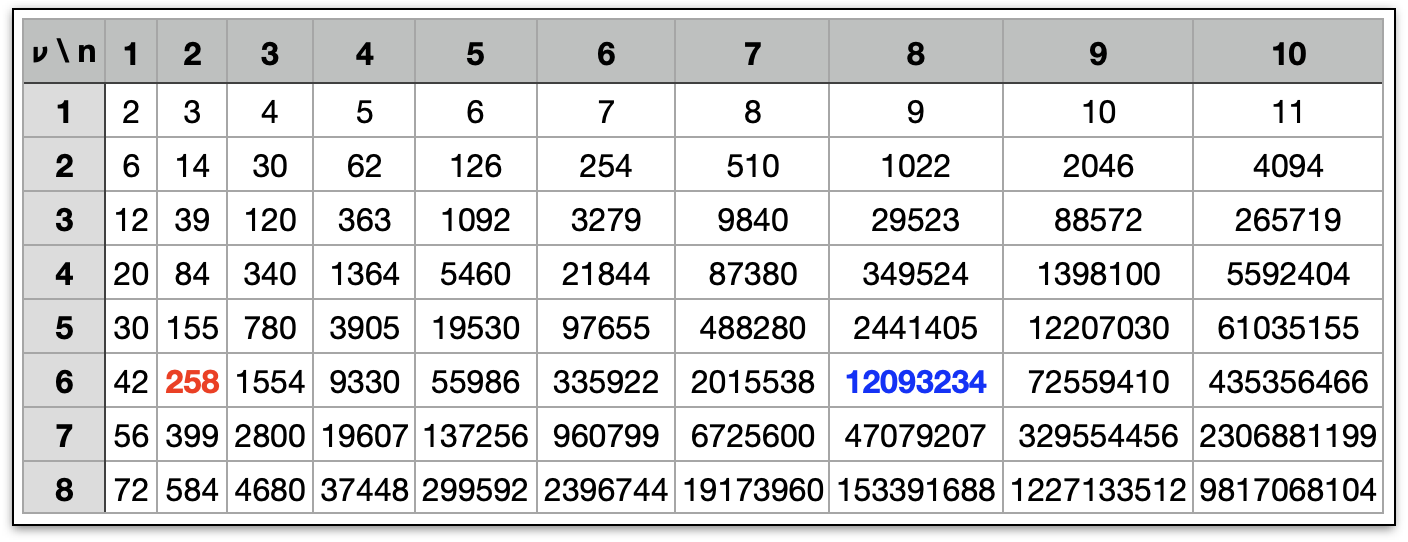}
\caption{Number of coefficients stored in DA maps as a function of the number of variables $\nu$ and orders $n$ using TSPA (top) and Matrix (bottom) representations.}
\label{tpsa_size}
\end{figure}

Figure~\ref{gtpsa_layout} illustrates the layout of such a DA map representing a phase space with 6 variables of order 2 and 3 parameters of order 1. Each \textit{variable} is a row-wise GTPSA that represents a truncated analytic function in the variables and parameters as multivariate Taylor polynomials. Each \textit{parameter} is defined as GTPSA of order 1, independently of the maximum order of the parameters in the variables set by \lstinline{po}, with their first derivative set according to their position in the parameters' list. Matrix codes like MAD-X use similar representation (without parameters) but are organized block-wise as illustrated by the orbit column vector $E$ (red), the 1st derivatives matrix $R$ (blue), and the 2nd derivatives tensor $T$ (pink). This matrix representation is not well suited for higher orders as shown in Figure~\ref{tpsa_size} where tensor sizes grow exponentially with numbers of variables and orders, making it difficult to carry out optics calculation even with 6 variables at order 5 with a size about 20 times bigger than TPSA. Keep in mind that the algorithmic complexity of polynomial multiplication scales quadratically with size!

\begin{figure}
\includegraphics[width=\linewidth]{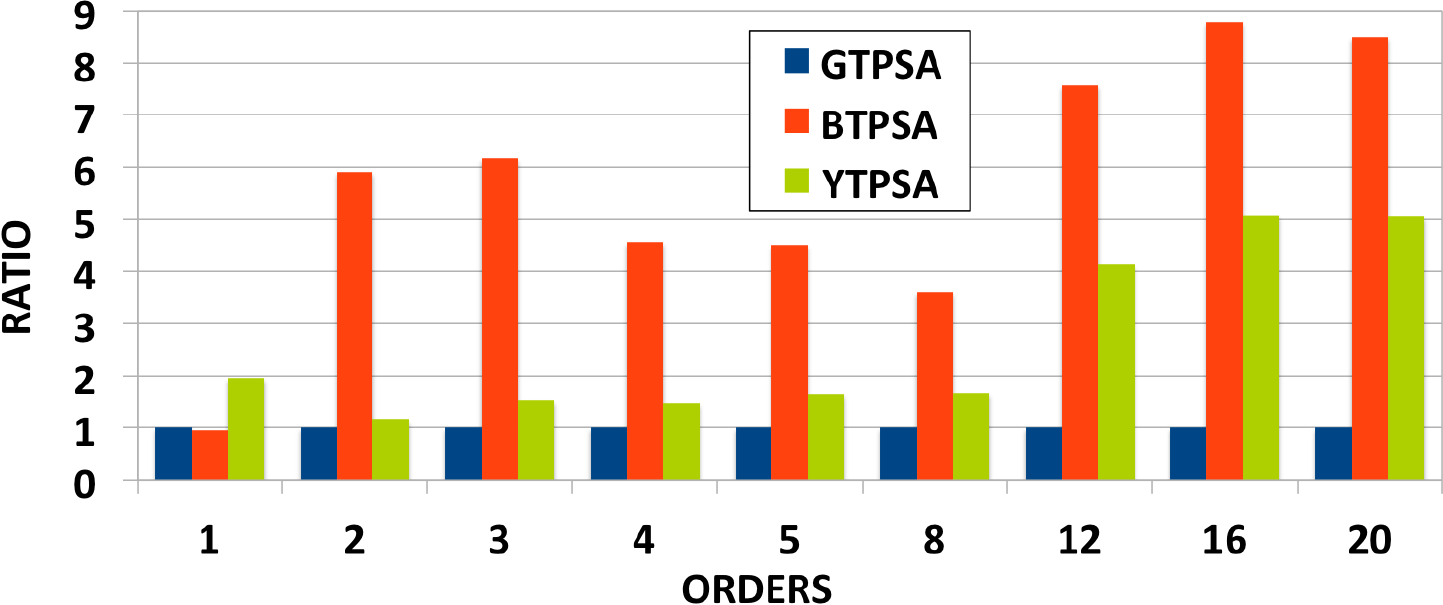} \\[1ex]
\includegraphics[width=\linewidth]{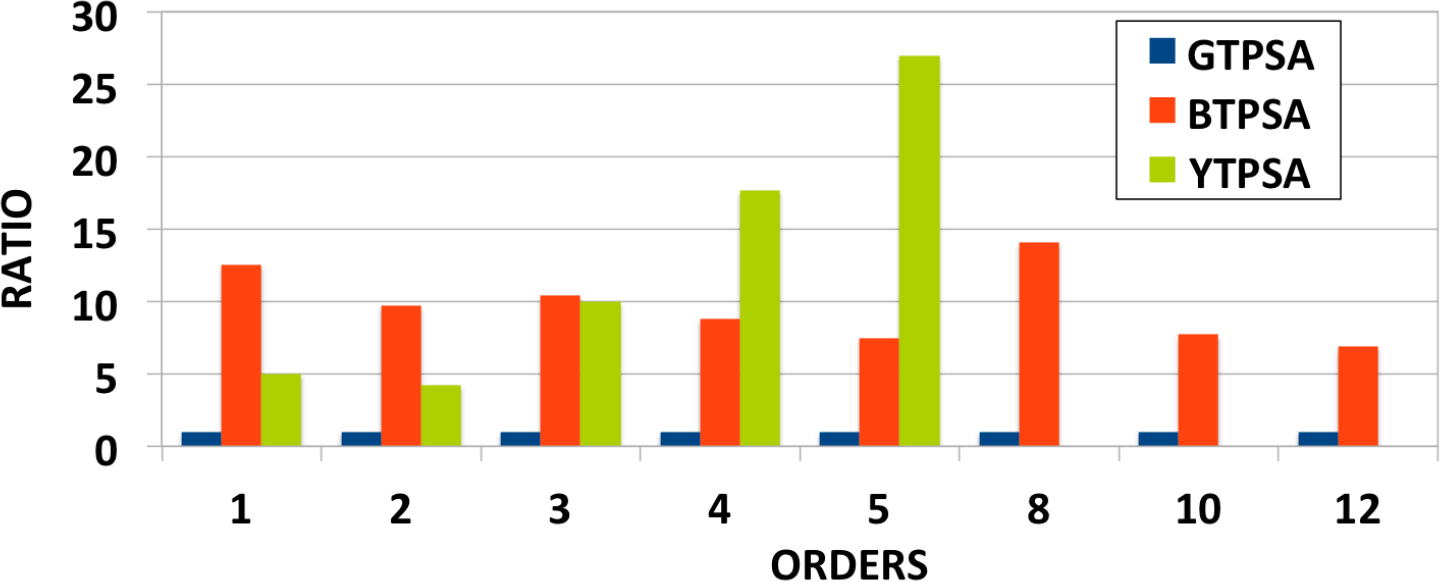} \\[1ex]
\includegraphics[width=\linewidth]{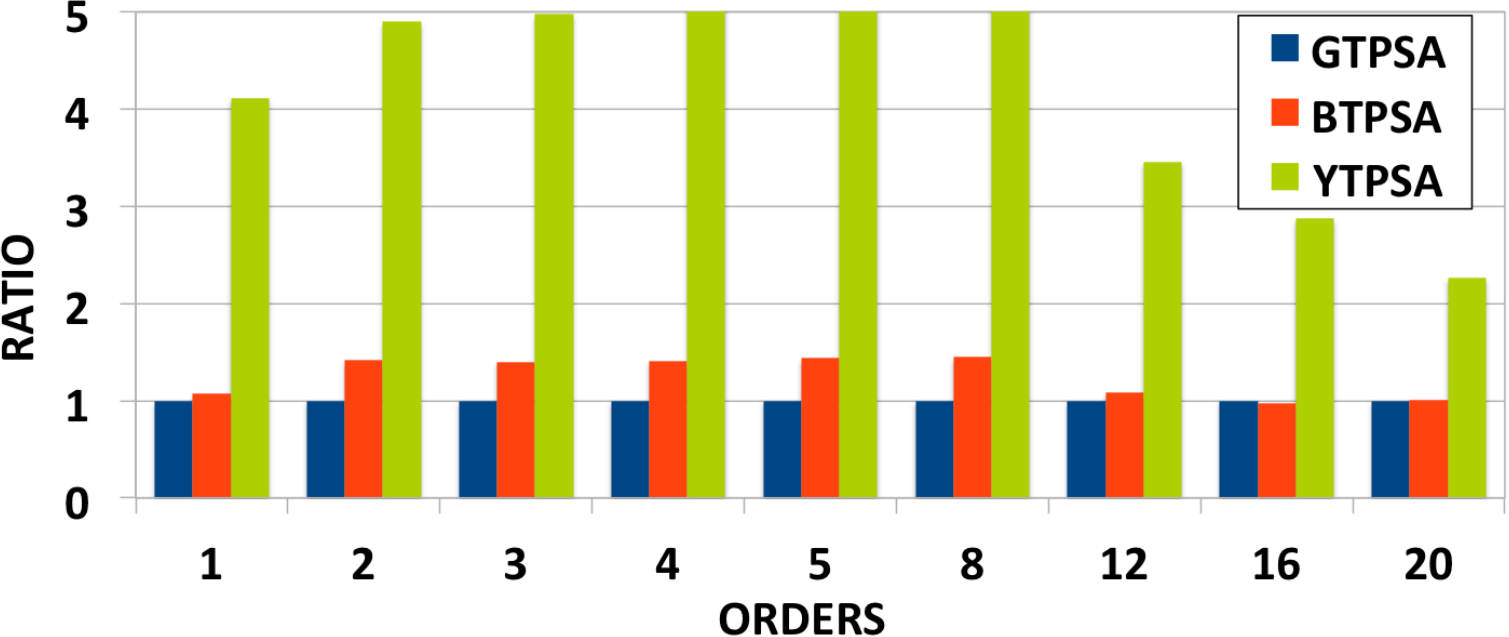}
\caption{Performance ratio of GTPSA vs. Berz's (BTPSA) and Lee Yang's (YTPSA) TPSA package for the TPSA multiplication (top), the DA maps composition (middle) and the conversion of monomial to index (bottom), larger is better in favor of GTPSA.}
\label{gtpsa_perfs}
\end{figure}

Considering that MADX-PTC already utilizes a TPSA framework developed by M.~Berz for high-order computations and can be setup to use L.~Yang (very limited) package as a replacement, why a new GTPSA package was needed? In fact, the Berz's package uses a simple and fast tensor-like indexation on sparse vectors, also known as the Kronecker trick for converting multivariate ($n$D) to single (1D) indexation. This indexation scheme can handle high orders but grows exponentially with the number of variables and parameters as already seen in Figure~\ref{tpsa_size}. For example, using 32 parameters of order 1 increases the indexing function range by $2^{32}$, meaning that even 64-bits integers indexes could only handle around 50 parameters. Therefore, a novel indexing function was invented to support hundreds of variables and parameters with non-uniform orders -- a capability uniquely offered by the GTPSA library, which forms the foundation of MAD-NG.

\begin{figure}
\includegraphics[width=\linewidth, angle=0]{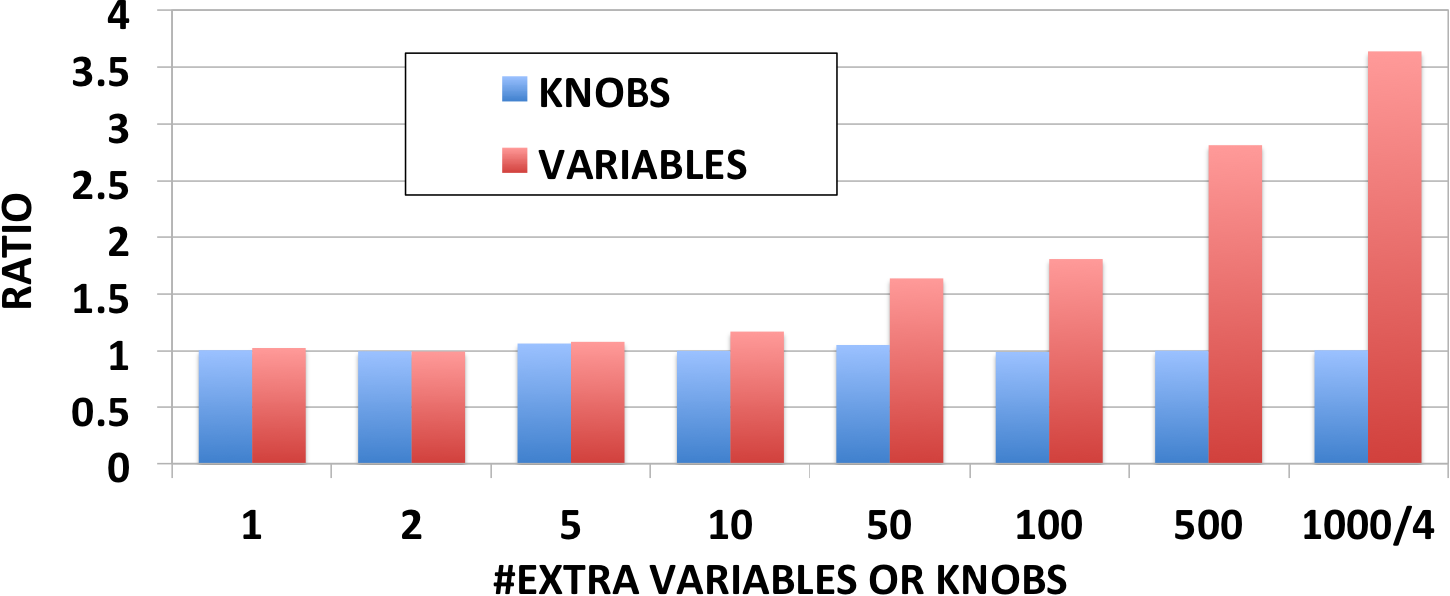}
\caption{Performance ratio of GTPSA with many parameters vs. plain TPSA with parameters treated as variables at order 2 (ratio for 1000 variables is divided by 4, ratio $\approx 15$).}
\label{gtpsa_knobs_perf}
\end{figure}

Figure~\ref{gtpsa_perfs} shows the performance of the GTPSA compared with M.~Berz's package (BTPSA) and L.~Yang's package (YTPSA), where the results are normalized by GTPSA, i.e. a ratio of 5 means that GTPSA are 5 times faster than the compared package. The three bar charts show the relative performance of the TPSA multiplication (top), DA maps composition (middle) and the indexing function (bottom) respectively. The indexing function converts a monomial into a vector index, working like a hash function.

Figure~\ref{gtpsa_knobs_perf} shows multiplication benchmarks demonstrating that GTPSA using parameters (blue) outperforms GTPSA using variables only (red) when handling a large number of 2nd order parameters. These results only apply the GTPSA library, as other packages are unable to handle such a large number of variables and parameters.

\section{Optimization with Match}

The \lstinline{match} command is an interface to MAD-NG's versatile optimizer offering many physics-oriented parameters to steer the results, and about 20 algorithms to tackle various kinds of problems.  

The following MAD-NG code snippet performs a sequence of operations to configure and match\footnote{MAD-NG's \lstinline{match} command follows the same "structure" as in MAD-X.} the LHC beam optics for beam 1 using the "relaxed" interaction point IP3 as the reference point 
to meet operational requirements in terms of global tunes and chromaticities.
\begin{lstlisting}
-- Load sequence and optics
MADX:load"lhc_as-built.seq"
MADX:load"opticsfile.1"
-- Attach a beam
local lhcb1, nrj in MADX
lhcb1.beam = beam {
  particle='proton', energy=nrj }
-- Set starting/observation point at IP3
lhcb1:cycle"IP3"
-- Match tunes and chromas
match {
  command := twiss {
    sequence=lhcb1, mapdef=2, observe=1},
  -- LHC B1 knobs
  variables = { rtol=1e-6, -- 1 ppm
    { name="dqx.b1" , var="MADX.dqx_b1"  },
    { name="dqy.b1" , var="MADX.dqy_b1"  },
    { name="dqpx.b1", var="MADX.dqpx_b1" },
    { name="dqpy.b1", var="MADX.dqpy_b1" },
  },
  -- Tunes and chromas target values
  equalities = { tol=2.5e-3,
    { name="q1" , expr=\t -> t.q1 -62.31 },
    { name="q2" , expr=\t -> t.q2 -60.32 },
    { name="dq1", expr=\t -> t.dq1-2     },
    { name="dq2", expr=\t -> t.dq2-2     },
  },
  -- Penalty function and options
  objective = { fmin=2e-3 },
  maxcall=100, info=2,
}
\end{lstlisting}
The script starts by loading the LHC sequence \lstinline{lhc_as-built.seq} and the corresponding optics settings file \lstinline{opticsfile.1}. These files define the layout, element properties, and initial optics configuration of the accelerator, including the beam energy \lstinline{nrj} used to attach to the sequence a new proton beam instance at the proper energy. The cycling stage moves the default starting point for tracking to a place where constrains are more relaxed than the default starting point IP1 which is the ATLAS's experiment place.

Follows the matching block that runs a \lstinline{twiss} command at order 2 several times -- one reference plus one per declared variable -- at each iteration of the optimizer to build the Jacobian matrix using adaptive finite differences. The Jacobian is then used by the optimizer to steer the tunes and the chromaticities returned in the \lstinline{twiss} table \lstinline{t} toward the targeted values specified in the \lstinline{equalities} constraints.

This can be achieved by acting on the LHC tuning knobs (variables) defined in the \MADX environment: \lstinline{dqx_b1} and \lstinline{dqy_b1} to control the horizontal and vertical tunes, and \lstinline{dqpx_b1} and \lstinline{dqpy_b1} to control the horizontal and vertical chromaticities. The relative tolerance \lstinline{rtol} for these variables is set to $10^{-6}$ (1 ppm), which is the power supplies relative accuracy. These knobs represent the main quadrupole and sextupole tuning "circuits" grouping a few hundred magnets through deferred expressions -- changing the value of these knobs affects the strengths of all these magnets at once.

\section{Radiation and Tapering}

Synchrotron radiation is the electromagnetic radiation emitted by charged particles (typically electrons or positrons) as they are accelerated through curved paths in a synchrotron or storage ring. In MAD-NG, particles lose energy from radiation in all elements that induce curvature in particles' motion, requiring turn-by-turn compensation via RF cavities to maintain their energy. Synchrotron radiation can be enabled with the \lstinline{radiate=true} attribute value -- other models like \lstinline{"damping"}, \lstinline{"quantum"}, \lstinline{"photon"}, and variants are available -- to the \lstinline{track}, \lstinline{cofind}, and \lstinline{twiss} commands. The latter enables only damping radiation and disables quantum radiation and photons tracking.

Tapering refers to the adjustment of the magnetic field strength along the length of a circular accelerator to compensate for energy losses between RF cavities sections. MAD-NG provides two ways to taper a lattice element by element, the simplest method being enabled with the \lstinline{taper=true} attribute value -- or the number of turns where tapering is active (fix point) -- to the \lstinline{track}, \lstinline{cofind}, and \lstinline{twiss} commands. The second method uses the \lstinline{taper} command that offers more options, and which enforces entering each element with zeroed transverse coordinates to avoid large orbit excursions on first passes. Both methods update the element's strength scaling factor \lstinline{ktap} to compensate the local energy losses, which is subsequently used to scale the strength $k_n (1+k_{\text{tap}})$ by the tracking engine. The \lstinline{taper} command also allows to save to and load from MAD tables the $k_{\text{tap}}$ values, giving users the opportunity to manipulate them easily and produce more realistic models, for example by averaging them along lattice's sections and magnet families to better model tapered circuits.

\begin{figure}
\includegraphics[width=1.01\linewidth]{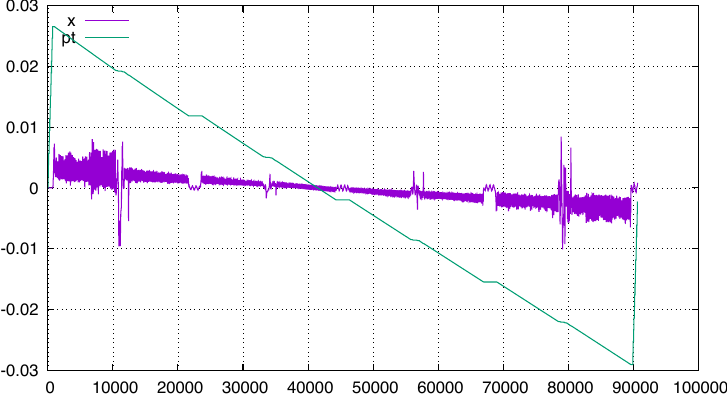} \\[2ex]
\includegraphics[width=\linewidth]{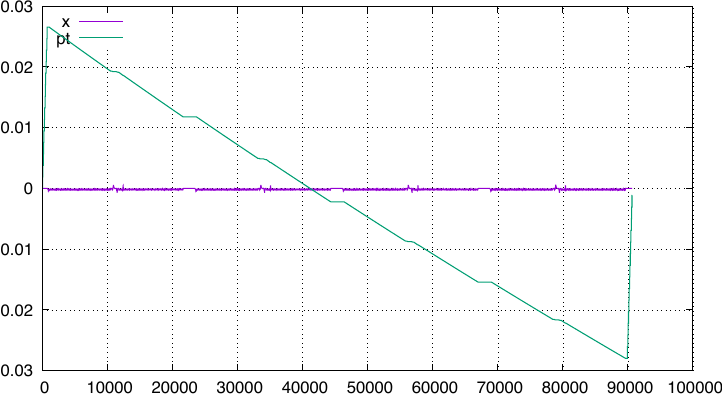} \\[2ex]
\hspace*{-2mm}
\includegraphics[width=1.02\linewidth]{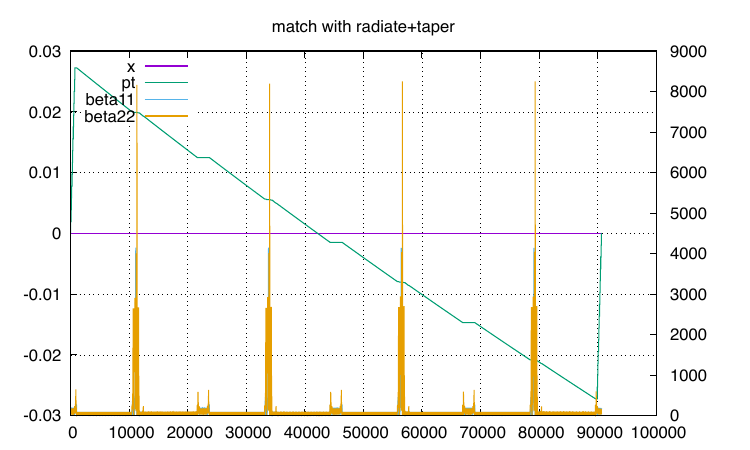}
\caption{Example of tapering the FCC-ee GHC optics at TTbar with radiation and RF cavities on. These plots show the $x, p_t$ coordinates along the ring without tapering (top), with a single pass tapering (mid), and with a multi-passes tapering combined with a closed orbit search during the matching of the RF cavities lags to balance the energy losses. All the axis are in meters except $p_t$ which is a normalized energy deviation.}
\label{fcc_taper}
\end{figure}

The following script example presents the use of the \lstinline{taper} command on the FCC-ee GHC lattice at TTbar (182.5 GeV) with a single section of RF cavities, split into two groups of 400~MHz and 800~MHz frequencies with unmatched initial lags in $2\pi$ unit \lstinline{lagca1 = 0.30} and \lstinline{lagca2 = 0.34} respectively.
\begin{lstlisting}
-- Load sequence and optics (single file)
MADX:load"fccee_t.seq"
-- Attach a beam
local fcc = MADX.fccee_p_ring
fcc.beam = beam {particle='positron',pc=182.5}
-- Cycle in the middle of the RF section
fcc:cycle"SP.8" :deselect(observed)
        ["SP.8"]:select  (observed)
-- Track with radiation (1st plot)
track {sequence=fcc, radiate=true, observe=0}
-- Taper the lattice
taper {sequence=fcc, taper=true}
-- Track with radiation (2nd plot)
track {sequence=fcc, radiate=true, observe=0}
-- Match RF cavities lags while tapering
match {
  command = function ()
    -- Taper the lattice
    taper {sequence=fcc, taper=true}
    -- Search for the closed orbit
    return cofind { -- Return Track MTable
      sequence=fcc, radiate=true, save=true} 
  end,
  -- RF cavities lags (400 MHz and 800 MHz) 
  variables = { rtol=1e-8, min=0.25, max=0.5,
    { name="lagca1", var="MADX.lagca1" },
    { name="lagca2", var="MADX.lagca2" },
  },
  -- Minimized energy deviation at "SP.8"
  equalities = { tol=1e-6,
    { name="pt", expr=\t -> t.pt[#t] },
  },
  -- Fine local tuning with bisections 
  objective = { bisec=5 },
}
-- Twiss with radiation (3rd plot)
twiss { sequence=fcc, radiate=true }
\end{lstlisting}

The script computes several results plotted in Figure~\ref{fcc_taper} with the \lstinline{plot} command (removed from example), but only the \lstinline{match} command section is relevant for tapering the lattice. The first result (top plot) shows the horizontal orbit deviation along the ring due to synchrotron radiation resulting to an energy deviation $p_t=0.002$ at the end of the tracking. The second result (middle plot) shows the effect of a single application of the \lstinline{taper} command on the horizontal orbit deviation along the ring while compensating the energy loss. The  energy deviation $p_t$ is reduced at the end, but the RF cavities are not yet matched meaning that the energy is not correctly adjusted, which is the purpose of the next step. The \lstinline{match} command section search the best lags that minimize the energy deviation $p_t$ using both, the \lstinline{taper} and the \lstinline{cofind} (closed orbit search) commands successively. After about 10 iterations, the following results summary is displayed by the optimizer:

{\scriptsize
\begin{verbatim}
Constraints  Type         Kind         Weight     Penalty Value
---------------------------------------------------------------
pt           equality     .            1          7.09330e-07
\end{verbatim}}

{\scriptsize
\begin{verbatim}
Variables    Final Value  Init. Value  Lower Limit  Upper Limit
---------------------------------------------------------------
lagca1       2.96787e-01  3.00416e-01  2.50000e-01  5.00000e-01
lagca2       3.34510e-01  3.42360e-01  2.50000e-01  5.00000e-01
\end{verbatim}}
with the final closed orbit coordinates $x=1.3\,10^{-9}$, $p_x=3.5\,10^{-12}$, $y=0$, $p_y=0$, $t=3.1\,10^{-6}$, $p_t=-7.1\,10^{-7}$. The final result (bottom plot) shows the check of the nearly perfect horizontal orbit deviation along the ring for the matched energy, as well as the horizontal and vertical optical beta functions (right axis) calculated along the ring by the \lstinline{twiss} command.

\section{Parametric Maps}

To ease and speedup optimization with many knobs as well as provide deeper insights into physics sensitivities, MAD-NG offers a unique feature called \textit{parametric differential maps} build from GTPSA, which combined with other well-designed features helps simplify the overall process:
\begin{enumerate}
\item Create a parametric phase-space and \textit{link} the knobs, e.g. magnet strengths, to the phase space parameters.
\item Optimize the constraints by varying the knobs using the derivatives of relevant quantities versus these knobs.
\item Restore the knobs as scalars with optimized values.
\end{enumerate}

Setting up parametric phase space in MAD-NG is simple, even when everything has been loaded inside the \MADX environment. For convenience, we start by defining the lists of variable and parameter names of the 6D parametric phase space such that they can be accessed uniformly by name everywhere:
\begin{lstlisting}[mathescape=true]
local prms = { -- Param./knob names (strings)
  -- 16 strengths of trim quadrupoles families
  "kqtf.a12b1","kqtf.a23b1",$\cdots$,"kqtf.a81b1",
  "kqtd.a12b1","kqtd.a23b1",$\cdots$,"kqtd.a81b1",
  -- 16 strengths of octupoles families
  "kof.a12b1" ,"kof.a23b1" ,$\cdots$,"kof.a81b1",
  "kod.a12b1" ,"kod.a23b1" ,$\cdots$,"kod.a81b1",
}
\end{lstlisting}
From these lists, we can define the parametric phase-space with \lstinline{nv=6} variables of order \lstinline{mo=5} and \lstinline{np=32} parameters of order \lstinline{po=1} named after the knobs in the optics file. The order of the \textit{variables} must be \lstinline{mo=4+1} as we want to optimize octupolar resonances at order $4$ using their derivatives versus the knobs from order $5$ in the example of the next sections.
\begin{lstlisting}
-- DA map representing parametric phase space
local X0 = damap {nv=6,mo=5,np=#prms,pn=prms}
\end{lstlisting}
The next step is to link the knobs defined in the optics file to the parameters defined in the phase space by replacing the knobs (scalars) in the \MADX environment with their corresponding parameters (GTPSA) from the phase space. Thanks to MAD-NG's deferred expressions and physics support for polymorphism, which simplifies the whole process through automatic handling by the \lstinline{track} and \lstinline{twiss} commands as already mentioned. This is achieved by the loop hereafter that replaces the knobs in \MADX by the sum of knobs and parameters with identical names such that initial conditions (e.g. injection optics) are preserved. 
\begin{lstlisting}
-- Convert scalars to GTPSAs within MADX env.
for _,knb in ipairs(prms) do
  MADX[knb] = MADX[knb] + X0[knb]
end
\end{lstlisting}
It is worth noting that any attribute of the lattice elements that might affect the constraints can be used transparently as DA maps parameters, not only the strengths and RF phases, but also lengths, positions, misalignments, etc might also be useful parameters for optimization or sensitivity studies. 

Once a suitable solution has been found by the optimizer (see next sections) using the parametric maps, we restore the knobs/strengths as scalars in the \MADX environment for further use.
\begin{lstlisting}
-- Restore knobs within MADX env. as scalars
for _,knb in ipairs(prms) do
  MADX[knb] = MADX[knb]:get0()
end
\end{lstlisting}

\section{Parametric Normal Forms}

\begin{figure*}
\centering
\includegraphics[width=\linewidth, angle=0]{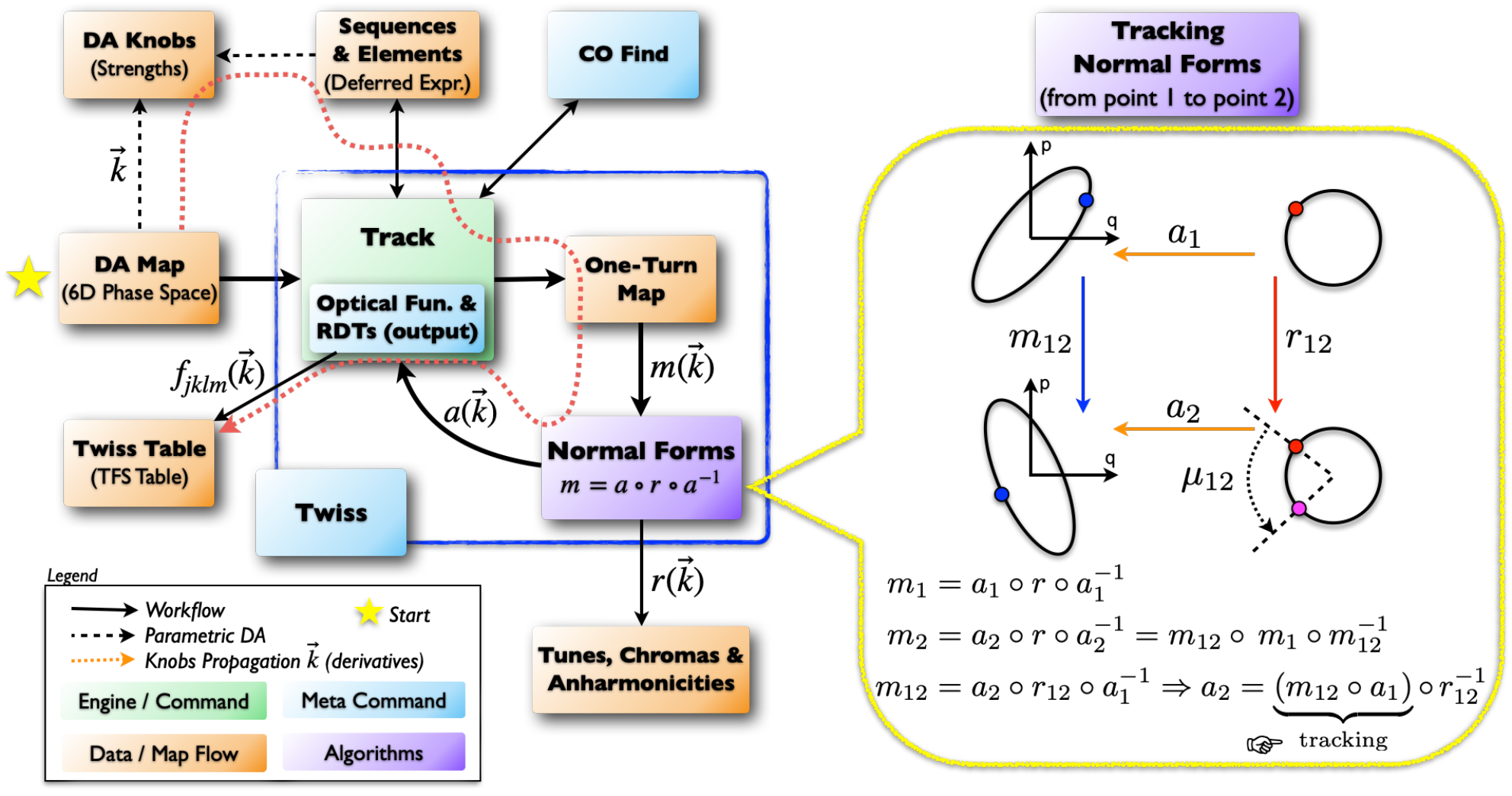}
\caption{The parametric normal form principle in \lstinline{twiss} consists of tracking a high-order DA map on the found closed orbit (optionally) equipped with parameters $\vec{k}$ (knobs) to obtain the one-turn map $m$, then compute the closed non-linear normal form $m=a\circ r \circ a^{-1}$ and track the normalizing map $a$ to extract the optical functions ($\alpha$, $\beta$, $\mu$, etc.) and the RDTs along the lattice.}
\label{normal-form}
\end{figure*}

To optimize, for instance, octupolar resonances, it is necessary to \textit{analyze the non-linear normal forms} of the parametric one-turn map generated by the \lstinline{track} command run on the design orbit (the lattice's reference trajectory), which is performed by the function \lstinline{get_nf} hereafter. Since, in this example, the objectives imply keeping the tunes and the amplitude detuning under control, we need to run \lstinline{get_nf} once to save these reference quantities before entering the optimization process.
\begin{lstlisting}
-- Function to compute non-linear normal forms
-- in a single point of the lattice
local function get_nf()
  local _, mflw = track {sequence=lhc, X0=X0}
  return normal(mflw[1]):analyse()
end
-- Save reference tunes and amplitude detuning
local nf     = get_nf()
local q1ref  = nf:q1{1}
local q2ref  = nf:q2{1}
local q1jref = nf:anhx{1,0}
local q2jref = nf:anhy{0,1}
\end{lstlisting}

\begin{figure}
\includegraphics[width=\linewidth]{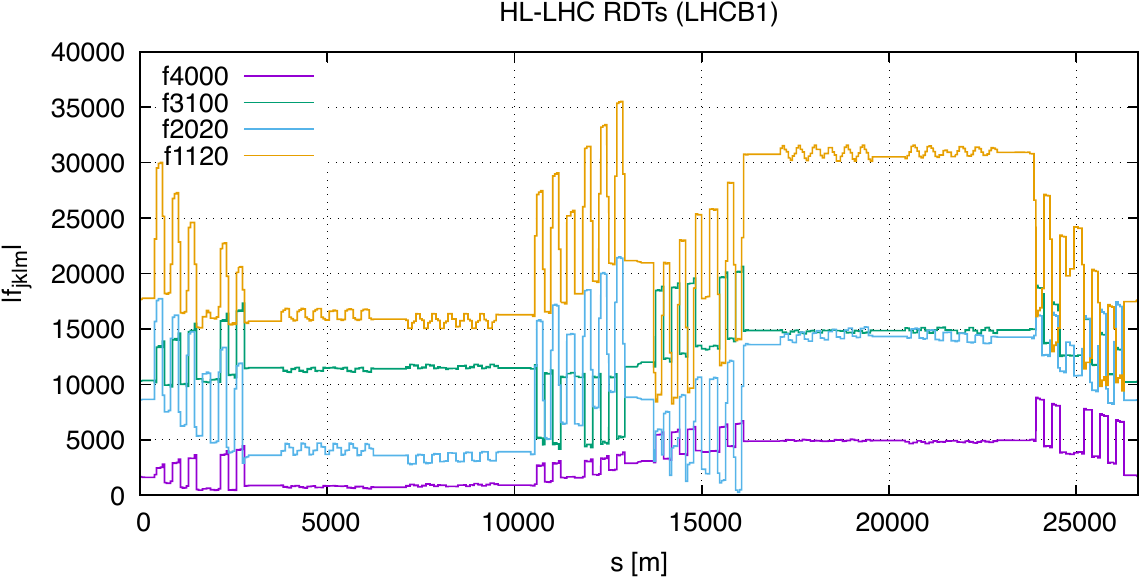} \\[2ex]
\includegraphics[width=\linewidth]{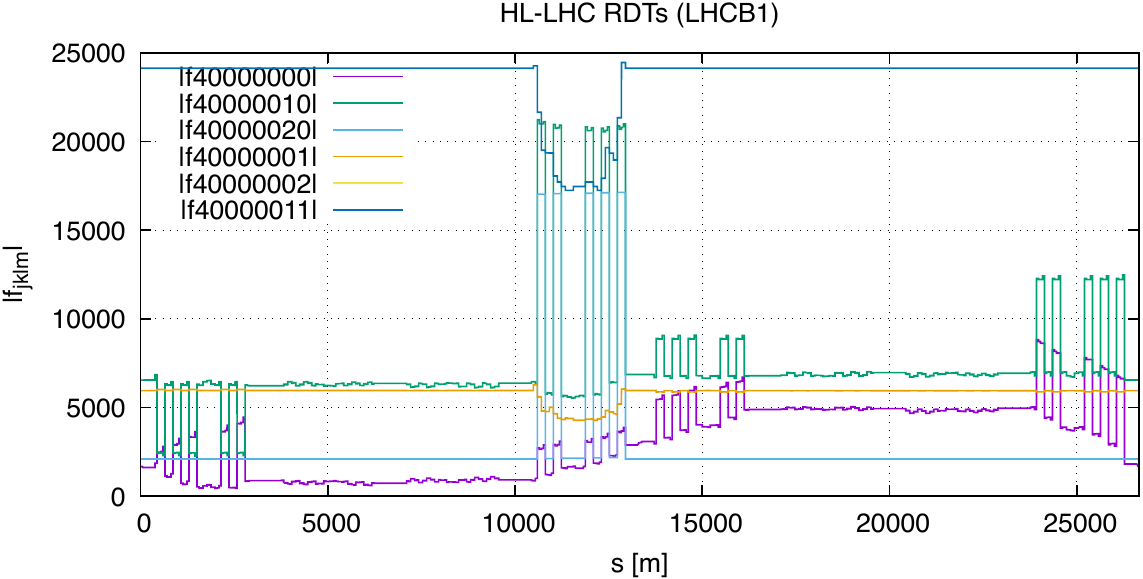}
\caption{HiLumi LHC octupolar RDTs (top) and the sensitivity of $4Q_x$ resonance versus the focusing (1st knob) and defocusing (2nd knob) sextupole strengths of the arc 45 (only) where $s\in[11,13]$ Km (bottom).}
\label{lhc_rdts}
\end{figure}

Another important use of the parametric normal forms includes sensitivity studies to resonances by analyzing for example the octupolar RDTs along the machine. For this purpose, the \lstinline{twiss} command is the simplest method as it will include RDTs calculation and extend its results table as soon as an RDTs list is provided via its \lstinline{trkrdt} attribute:
\begin{lstlisting}
-- List of RDTs
local rdts = {"f4000","f3100","f2020","f1120"}
-- Create phase-space damap at 4th order
local X0 = damap {nv=6, mo=4}
-- Compute RDTs along HL-LHC
local tw = twiss {sequence=lhcb1, X0=X0,
                  trkrdt=rdts}
\end{lstlisting}

The complete process triggered by linear and non-linear normal form calculation, tracking, and analysis is explained in the key Figure~\ref{normal-form}. The RDTs available as extended output from the \lstinline{twiss} table \lstinline{tw} can be plotted along the machine as in Figure~\ref{lhc_rdts} (top) and for visual analysis and understanding of local or global perturbative effects (bottom). Compared to the former approach (top), the latter approach (bottom) will use a parametric DA map with parameters at order 2 to obtain quadratic and cross-term sensitivity:
\begin{lstlisting}
-- List of knobs
local prms = {"ksf1.a45b1","ksf2.a45b1"}
-- List of RDTs
local rdts = {"f40000000", "f40000010",
              "f40000001", "f40000011", 
              "f40000020", "f40000002"}
-- Create a 6th order damap
local X0 = damap {nv=6, mo=6, np=#prms,
                        po=2, pn= prms}
-- Compute RDTs sensitivity along HL-LHC
local tw = twiss {sequence=lhcb1, X0=X0,
                  trkrdt=rdts}
\end{lstlisting}
The calculations of the two previous snippets and the plots of Figure~\ref{lhc_rdts} take 16 seconds and around 1 minute for the top and bottom plots respectively on my laptop, making these analyses more accessible compared to what was previously achievable without MAD-NG. 

The algorithm for the calculation of the non-linear normal forms can be found in\cite{Forest15,Forest89}. MAD-NG's parametric non-linear normal forms and analysis implementations are variants based on the Lie operators of the GTPSA library. The comparison of MAD-NG versus MADX-PTC has proven to give the same results for RDTs calculation in many studies on several CERN accelerators, but with an improved calculation speed ranging from $\times 30$ to $\times 80$ faster. 

\section{Parametric Optimisation}

The main difference between the optimization with finite differences and parametric maps is the order of the DA map representing the phase-space, which is one order higher than the non-parametric version to extract the \textit{exact} Jacobian. The downside of finite differences is that for each iteration of the optimizer, the \lstinline{get_nf} function has to be invoked 1+32 times to \textit{approximate} the Jacobian with finite differences, whereas with parametric maps the function \lstinline{get_nf} has to be run only once. If we consider the map sizes ratio of previous section, which is between order 4 with no parameter and order 5 with 32 parameters to be $1260\times33/(2772+1260\times32)\approx 0.97$, and a computational time almost linear in size, the two methods should be equivalent in speed. But in practice parametric maps appear to be three times faster on average, and converge faster to the solution as they do not rely on fragile estimates of finite differences steps.

The following script shows the \lstinline{match} command being executed in an attempt to satisfy constraints specified as \textit{equalities} and expressed as lambda functions returning values to be compared with zero, i.e. the targeted/reference values must be subtracted. The optimizer will vary the values of the \textit{variables} according to results returned by the \textit{command} invoked once (or 33 times for finite differences) for each iteration.
\begin{lstlisting}[mathescape=true]
match {
-- Compute non-linear normal forms
command := get_nf(), -- returns nf used below
-- Compute Jacobian from parametric maps
jacobian = \nf,_,J =>
  for k=1,32 do -- fill [10x32] J matrix                  
    J:set(1,k, nf:q1{1,k} or 0)
    J:set(2,k, nf:q2{1,k} or 0)
    J:set(3,k, nf:anhx{1,0,0,k})
    J:set(4,k, nf:anhy{0,1,0,k})
    J:set(5,k, nf:gnfu{"2002",k}.re)
    J:set(6,k, nf:gnfu{"2002",k}.im)
    J:set(7,k, nf:gnfu{"4000",k}.re)
    J:set(8,k, nf:gnfu{"4000",k}.im)
    J:set(9,k, nf:gnfu{"0040",k}.re)
    J:set(10,k,nf:gnfu{"0040",k}.im)
  end
end,
-- Variables in MADX env. to use as knobs
variables = {
  {name=prms[1] , var='MADX[prms[1]]' },
  {name=prms[2] , var='MADX[prms[2]]' },
     $\cdots$
  {name=prms[32], var='MADX[prms[32]]'},
},
-- Target constraints as equalities to zero
equalities = {
  {name='q1'  ,expr=\nf->nf:q1{1}-q1ref},
  {name='q2'  ,expr=\nf->nf:q2{1}-q2ref},
  {name='q1j1',expr=\nf->nf:anhx{1,0}-q1jref},
  {name='q2j2',expr=\nf->nf:anhy{0,1}-q2jref},
  {name='f2002r',expr=\nf->nf:gnfu"2002".re},
  {name='f2002i',expr=\nf->nf:gnfu"2002".im)},
  {name='f4000r',expr=\nf->nf:gnfu"4000".re)},
  {name='f4000i',expr=\nf->nf:gnfu"4000".im)},
  {name='f0040r',expr=\nf->nf:gnfu"0040".re)},
  {name='f0040i',expr=\nf->nf:gnfu"0040".im)},
}, 
} -- close match
\end{lstlisting}
The key point is the function \lstinline{jacobian} which is called once (if present) in place of the 33 calls to obtain the Jacobian \lstinline{J} required by the optimizer. It queries the analyzed normal forms with the extra argument \lstinline{k} representing the index of the parameter, e.g. \lstinline|nf:gnfu{"2002",k}| returns the complex value $\partial f_{2002} / \partial K_k$ from fifth order, where $K_k$ is the knob (magnet strength) associated with the $k$-th parameter of the phase-space.

The reference values are computed in less than 3~s on my laptop. Full optimization is achieved in 21 evaluations of \lstinline{get_nf} and 65~s with MAD-NG. Previously the same study using MADX-PTC was taking 342 evaluations and 2730~s to converge, i.e. being about $42$ times slower.

\begin{figure}
\begin{center}
\includegraphics[width=\linewidth,trim=0 0 0 45,clip]{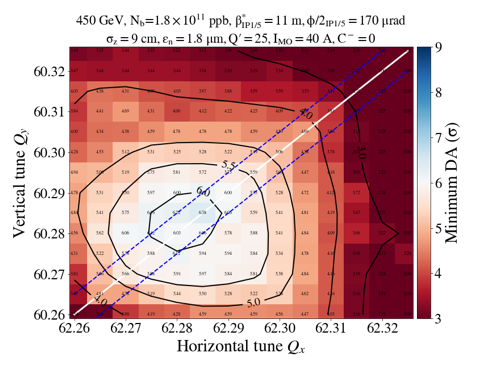} \\
\includegraphics[width=\linewidth,trim=0 0 0 45,clip]{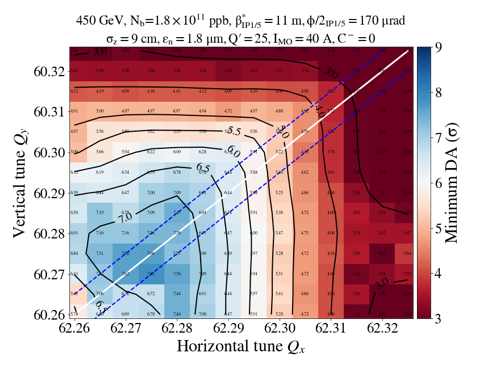}
\caption{Dynamic aperture for beam 1 old (top) and new (bottom) LHC injection optics (beam 2 has similar results).}
\label{newDA}
\end{center}
\end{figure}

The primary figure of merit for an optics design in the non-linear regime is the Dynamic Aperture (DA), defined as the minimum transverse amplitude beyond which particles become unstable. Figure~\ref{newDA} compares the DA for the old (top) and new (bottom) optics as a function of horizontal and vertical tunes, highlighting the improved performance of the new optics with optimized octupolar resonances\footnote{These plots were made with another tool by S. Kostoglou at CERN.}. Operational experience with the new injection optics, detailed in~\cite{tomasHB}, confirms enhanced beam lifetime of the LHC.

\section{Conclusions}

MAD-NG has successfully completed its development phase, with version 1.0 released at the end of 2024. The tool is now widely used in various CERN studies and lattice optimization tasks, demonstrating its accuracy and efficiency in solving complex linear and non-linear problems.

We demonstrated that the parametric normal form is a powerful tool for analyzing the sensitivity of various quantities to parameters such as strengths, lengths, positions, misalignments, and cross-talk, providing valuable insights into accelerator behavior or inputs to the optimizer.

MAD-NG offers several key advantages over MAD-X, including a more robust structure and significantly improved physics capabilities inherited from PTC/FPP. Its highly flexible and extensible architecture allows new features to be implemented within days. Additionally, it does not trig unexpected behaviors when combining complex features together like slicing, misalignments, frame patches, field errors, and combined functions.

MAD-NG also introduces support for advanced functionalities such as backtracking, charged particles, parallel sequences, and reversed sequences. These capabilities open up new possibilities for studies, which the MAD-NG community is beginning to explore and integrate into their workflows.

Moreover, MAD-NG uses a mainstream programming language for scripting and an outstanding Just-in-time compiler, significantly reducing user time with its mature technology, robust syntax error handling, inline debugging, and comprehensive backtracing features.

MAD-NG was developed to \textit{methodically} design new particle accelerators, where it is expected to excel in the near future.

%%%%%%%%%%%%%%%%%%%%%%%%%%%

\end{document}